\documentclass{article}

\usepackage[final]{neurips_2024}

\usepackage[utf8]{inputenc} 
\usepackage[T1]{fontenc}    
\usepackage[backend=biber, style=numeric-comp, sorting=none, doi=true, eprint=true, maxbibnames=999]{biblatex}
\usepackage{hyperref}
\hypersetup{
  colorlinks=true,
  urlcolor=brown,
  citecolor=blue
}

\DeclareFieldFormat{eprint:arxiv}{%
  arXiv:\href{https://arxiv.org/abs/#1}{#1}%
}

\DeclareFieldFormat{doi}{%
  \ifhyperref
    {\href{https://doi.org/#1}{\nolinkurl{https://doi.org/#1}}}
    {\nolinkurl{https://doi.org/#1}}}

\usepackage{url}            
\usepackage{booktabs}       
\usepackage{amsfonts}       
\usepackage{nicefrac}       
\usepackage{microtype}      
\usepackage{xcolor}         

\usepackage[pdftex]{graphicx}
\usepackage{amsmath}
\usepackage{amsfonts} 
\usepackage{amsthm}   
\usepackage{amssymb}  
\usepackage{subcaption} 
\usepackage{tikz} 
\usepackage[normalem]{ulem}

\newcommand\bra[2][]{#1\langle {#2} #1\rvert}
\newcommand\ket[2][]{#1\lvert {#2} #1\rangle}

\newcommand{\tr}{\textrm{Tr}}

\usepackage{comment}

\title{Universal quantum phase classification on quantum computers from machine learning}

\author{%
  Weicheng Ye\\
  Department of Physics and Astronomy, and Stewart Blusson Quantum Matter Institute,\\
  University of British Columbia, Vancouver, BC V6T 1Z1, Canada\\
  \texttt{victoryeofphysics@gmail.com} \\
  \And
  Shuwei Liu \\
  Perimeter Institute for Theoretical Physics, Waterloo, Ontario N2L 2Y5, Canada \\
  \AND
  Shiyu Zhou \\
  Perimeter Institute for Theoretical Physics, Waterloo, Ontario N2L 2Y5, Canada \\
  \And
  Yijian Zou \\
  Perimeter Institute for Theoretical Physics, Waterloo, Ontario N2L 2Y5, Canada \\
  \texttt{yzou@perimeterinstitute.ca} \\
}

\begin{document}

\maketitle

\begin{abstract}
The classification of quantum phases of matter remains a fundamental challenge in condensed matter physics. We present a novel framework that combines shadow tomography with modern time-series machine learning models to enable efficient and practical quantum phase classification. Our approach leverages the definition of quantum phases based on connectivity through finite-depth local unitary circuits, generating abundant training data by applying Haar random evolution to representative quantum states for a given phase. In this way, the training data can be efficiently obtained from a quantum simulator. Additionally, we demonstrate that advanced time-series models can be used to process the training data and achieve universal quantum phase classification that does not rely on local order parameters. To validate the universality and versatility of our method, we test the model against one-dimensional quantum spin chains such as the Ising model and the axial next-nearest-neighbor Ising (ANNNI) model, showing excellent agreement with known phase boundaries. 
\end{abstract}

\section{Introduction}

The classification of quantum phases of matter represents one of the most fundamental challenges in condensed matter physics \cite{sachdev2023quantum}, with profound implications for our understanding of emergent phenomena and quantum many-body systems. Traditional approaches to phase classification often rely on symmetry breaking with local order parameters, which can be insufficient for characterizing exotic phases such as topological states or spin liquids. Thus, there is a surge of interest in designing physical quantities, especially entanglement-based quantities, to identify and differentiate quantum phases,  ranging from topological entanglement entropy \cite{2006PhRvL..96k0404K, 2006PhRvL..96k0405L} to more recent proposals using multipartite entanglement \cite{Pezz__2016, Pezz__2017,2022PhRvL.129z0402Z,Fan2022,Siva_2022,Liu_2024}. A key feature of these quantities is that they are defined on a local \emph{patch} consisting of a finite number of qubits in a local region. While these approaches can be justified with additional assumptions \cite{Shi_2020,li2025strictarealawentanglement}, there are some limitations in practice. In particular, these quantities are usually not robust under finite-depth local unitaries (FDLU) which define equivalence relations of quantum phases \cite{Zou_2016,Williamson_2019,2023PhRvL.131p6601K,gass2024manybodysystemsspuriousmodular,Levin2024}. Thus, finding a substitute for them which is genuinely invariant under FDLU has been an active research area. Additionally, entanglement quantities are not direct observables on a quantum computer, and accessing these quantities typically requires full tomography on the reduced density matrix of a patch, which scales exponentially with the patch size \cite{Haah_2017,Yuen_2023}. It is therefore desirable to find an alternative way to probe quantum phases directly from measurement outcomes of a quantum computer. These measurement outcomes in randomized basis constitute the so-called \emph{shadow data} \cite{Huang_2020}, which are easy to obtain from quantum hardware. Our goal is to probe the quantum phases directly from these shadow data in a local patch, preferably with a smaller number compared with those required in full tomography. 

In recent years, the application of machine learning techniques to quantum phase classification has emerged as a promising avenue that offers new perspectives on this longstanding problem \cite{Carrasquilla_2017, van_Nieuwenburg_2017, Wang_2016, Wetzel_2017, Huang_2022,Cong_2019,Pollman2023,2024arXiv240211022C,bermejo2024quantumconvolutionalneuralnetworks,le2025distinguishing,2017PhRvB..96x5119Z,SanchoLorente_2022,An2024learningquantum,khosrojerdi2024learningclassifyquantumphases,2025npjQI..11...42F, 2017NatSR...7.8823B,2018FrPhy..13.0507W,2018PhRvB..98h5402S,2024PhRvA.109e2623F}, since it can potentially identify subtle patterns and correlations in quantum data that may not be immediately apparent through conventional analytical methods. These methods achieved impressive accuracy in detecting well-known transitions such as the Ising ferromagnetic transition \cite{Carrasquilla_2017} and have been extended to more complex scenarios including topological phases \cite{2017PhRvB..96x5119Z} and many-body localization \cite{2025npjQI..11...42F}. However, a main feature of these approaches is that the generation of training data often assumes good physical understanding and efficient simulation of the specific model, including Monte Carlo sampling or tensor network simulation, which hinders their generalization to other models where physical understanding is lacking. Recent progress, such as Ref.~\cite{le2025distinguishing}, first proposes to use experimental data from quantum simulations to learn quantum phases of matter, also utilizing system-specific features. These limitations highlight the need for more systematic, generalizable, and experimentally feasible approaches to machine learning-based phase classification.

The advent of modern time-series machine learning models, which have seen tremendous success in many different fields, presents a particularly valuable opportunity in this context.  There is already a lot of work on representing the ground state of physical Hamiltonian using modern time-series models such as recurrent neural network (RNN) \cite{2020PhRvR...2b3358H} or transformer \cite{fitzek2024rydberggpt,2025arXiv250702644G}. For classifying phases from shadow data, time-series models appear more intuitive because it is natural to think of the shadow data coming out of a \emph{sequence} of measurements performed in a certain order. Hence, we anticipate that modern time-series models, including RNN, convolutional neural network (CNN), and transformer are natural candidates for our classification task through capturing quantum correlations across the patch.\footnote{It is worth emphasizing that, in our work, the "time sequence" refers to a sequence of spatially distinct measurements obtained from a patch, rather than measurements at different time slices.} 

In this work, we address these challenges by constructing a systematic framework for quantum phase classification based on measurement data from quantum devices and modern time-series machine learning models. 
There are three key ingredients for our framework. Firstly, the input features are obtained from shadow tomography \cite{Huang_2020,elben2020mixed}, a technique which enables an efficient estimation of multiple properties of quantum states from a small number of measurements. Identifying quantum phases of matter with classical shadows is provably efficient with machine learning \cite{Huang_2022}. Secondly, we use the definition of quantum phases of matter in terms of FDLU circuits \cite{2010PhRvB..82o5138C}. This allows us to generate a large dataset with their labels determined solely from the definition. This approach in our task using classical machine learning models resembles the data generation for quantum machine learning in Ref.~\cite{Pollman2023} without symmetry constraints. In this way, the data generation can be achieved systematically without requiring analytical understanding or numerical simulations on many different lattice models. Moreover, our classification of phases does not rely on specific order parameters; instead, the local information is effectively scrambled by the random unitaries and then extracted by the machine learning model, circumventing limitations associated with order parameter-based methods. 
Thirdly, we utilize time-series machine learning models, such as recurrent neural networks (RNNs) and convolution neural networks (CNNs), as the machine learning model to classify quantum phases based on shadow data. These architectures are naturally suited for quantum many-body systems since they recognize patterns of measurement outcomes across different spatial locations. 

We illustrate our approach with a neural network classifier that distinguishes spontaneous symmetry broken (SSB) and trivial states in one dimensional quantum many-body states. We generate many quantum states for our training through \emph{Haar random} FDLU evolution acting on Greenberger-Horne-Zeilinger (GHZ) states and product states, which give enough representatives for SSB and trivial phases, respectively. The protocol for generating train dataset is discussed in Section~\ref{sec:haar_random}. We then employ classical shadow tomography to obtain measurement data from these evolved states as our training data. We feed these training data to an RNN or CNN based classifier, which is detailed in Section~\ref{sec:nn} and Appendix~\ref{app:CNN}. In Section~\ref{sec:ising}, we test our model on the physical Ising model and a generalization of it, axial next-nearest neighbor Ising (ANNNI) model. The ground states of our physical model do not appear explicitly in the training set, yet our machine learning model can still reproduce the known phase diagram, demonstrating the universality of our framework. We also compare our machine-learning model with two non-machine-learning quantities in Appendix~\ref{app:purity}, further highlighting the advantage of our machine learning framework. 

\section{Data generation from Haar random unitary evolution}\label{sec:haar_random}

The quantum phases we consider are defined on systems of $N$ qubits/spins, where $N$ is large. These systems can be efficiently simulated on modern quantum computers. Every qubit has a 2-dimensional Hilbert space $\mathbb{C}^2$, with basis states denoted by $\ket{1}$ and $\ket{0}$. An $N$-qubit \emph{quantum state} resides in the $2^N$-dimensional Hilbert space $\mathbb{C}^{2^N}$. Operators on this space, or so-called \emph{quantum gates}, are defined as a unitary operator acting on the Hilbert space, and form the group $U(2^N)$.

\begin{figure}[!htbp]
    \centering
\begin{subfigure}{\linewidth}
  \centering
  \includegraphics[width=\linewidth]{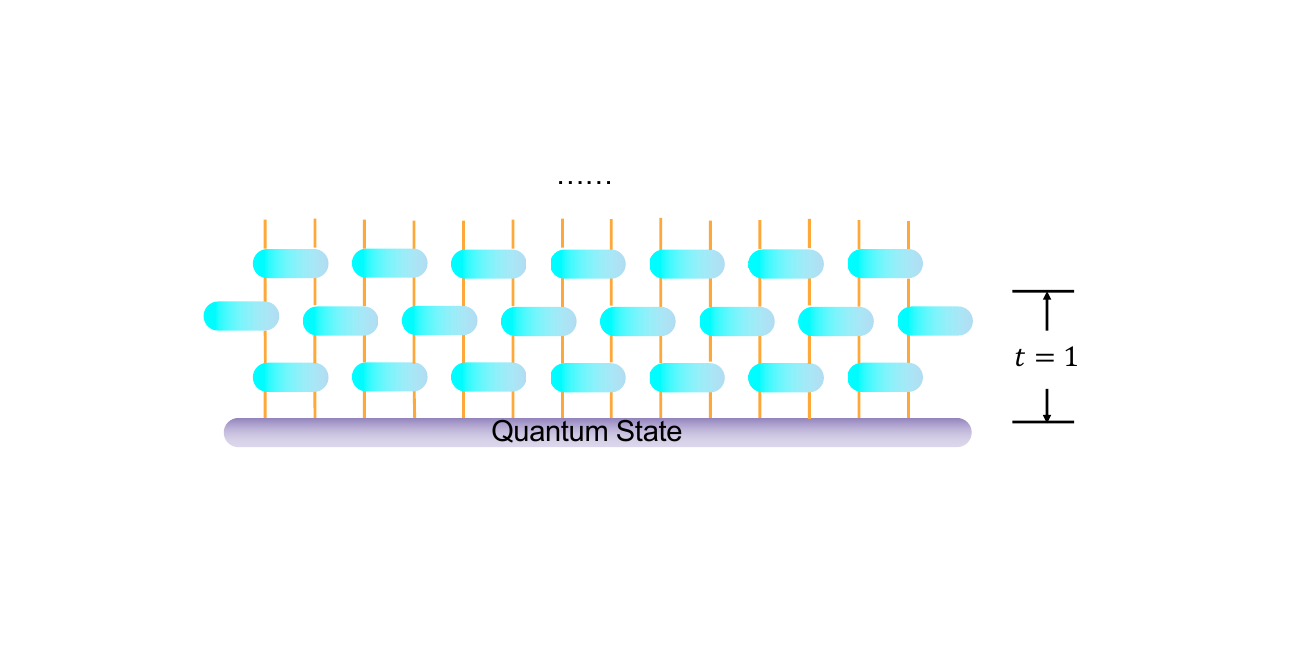}
  \caption{Finite depth local unitary evolution}
\end{subfigure}
\begin{subfigure}{\linewidth}
  \centering
  \includegraphics[width=\linewidth]{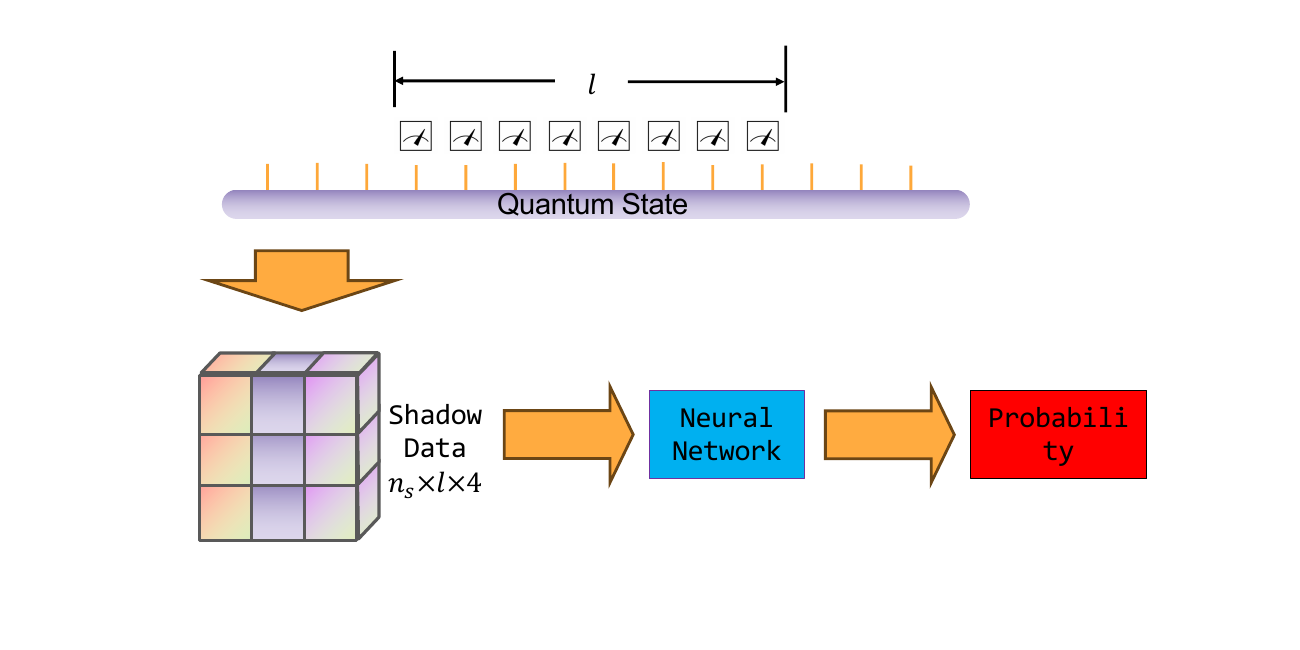}
  \caption{Protocol}
\end{subfigure}
    \caption{
        \textbf{Data Generation.} This schematic illustrates our data generation protocol based on finite-depth local unitary (FDLU) evolution. (1) \textbf{Finite-Depth Local Unitary (FDLU) Evolution.} The finite-depth local unitary evolution is defined as a series of two-qubit unitary gates applied sequentially to adjacent pairs of qubits, forming a brick-wall circuit structure. The depth of this circuit, denoted by $t$, determines the extent of local entanglement generated. By varying these random unitaries in a Haar random distribution across multiple iterations, we generate an ensemble of diverse states that belong to the same quantum phase as the initial representative state. (2) \textbf{Protocol.} This figure illustrates the protocol we use to do the classification. Given a quantum state, we perform single-site measurements on sites in a patch of size $l$ in a random basis and repeat $n_s$ times. This generates a tensor of dimension $n_s\times l \times 4$ for a single quantum state, which is the \emph{shadow data}. We then feed the obtained shadow data into a neural network and obtain the probability of the quantum state being in a given phase. 
    }
    \label{fig:haar}
\end{figure}

Our approach relies on the equivalence of \emph{quantum phases} based on \emph{finite-depth local-unitary} (FDLU) evolution \cite{2010PhRvB..82o5138C}. Namely, two quantum states $|\psi_0\rangle$ and $|\psi_1\rangle$ are defined to be in the same \textit{phase} if there is a finite-depth local unitary $U$ (as illustrated in Figure~\ref{fig:haar}(a)), with depth $t\leq \mathrm{polylog}(N)$, such that
\begin{equation}
    |\psi_1\rangle = U|\psi_0\rangle
\end{equation}
up to exponentially small corrections. This definition captures the essential topological nature of quantum phases and provides a rigorous framework for phase classification that encompasses both conventional symmetry-breaking phases and exotic topological phases that lack local order parameters. Our goal is to identify the quantum phase that a given quantum state belongs to from the information in a \emph{patch} consisting of $l$ adjacent qubits where $l\ll N$ .\footnote{The information on a patch is expected to contain all universal information of a gapped phase. The phases we consider, i.e. trivial and SSB phases, are all gapped phases.}

From the data-driven point of view, this definition naturally suggests a data generation protocol for phase classification. For a given quantum phase, we can start with known representative states of the phase, and generate a diverse set of states in the same phase through the action of Haar random FDLU evolution, as illustrated in Figure~\ref{fig:haar}(a). In the Haar random FDLU evolution, every block in Figure~\ref{fig:haar}(a) is a (Haar) random unitary in the set of unitary group $U(2^2)$ acting on two qubits. These randomly evolved states naturally explore the full breadth of the space of states for a given phase, providing a rich dataset that captures the intrinsic variability within each phase while being unbiased toward a specific basis or a specific order parameter. We then perform random measurements on a patch of $l$ spins and collect the measurement outcomes as our training data. The resulting data thus encodes the \textit{universal} properties of the phase. We note that the number of shadows has to grow at least super-polynomially with the depth $t$, due to a complexity theory bound put forward by Ref.~\cite{schuster2025randomunitariesextremelylow}. Thus, we will restrict the depth of the unitary circuit to be $t=O(1)$. More specifically, we will focus on $t=1$ and show that it is already practically useful for phase classification.

Concretely, the simplest possible state constructed out of the qubits is the \textit{product state}, e.g.,
\begin{equation}\label{eq:product}
   \otimes \left(\frac{1}{\sqrt{2}}\left(\ket{1} + \ket{0}\right)\right),
\end{equation}
where the total state of the spin chain is simply the tensor product of the states of individual spins. Another important class of state is called the \textit{Greenberger–Horne–Zeilinger (GHZ) state},
\begin{equation}\label{eq:GHZ}
    \frac{1}{\sqrt{2}}\left(|1 1 .. 1 \rangle + |0 0 .. 0 \rangle\right),
\end{equation}
which is a global superposition of two different product states. The two states cannot be connected to each other by FDLU due to the Lieb-Robinson bound \cite{_Anthony_Chen_2023}, thus they belong to different phases. 

Physically, these two states can be thought of as the ground states of the transverse field Ising model at two extreme limits. Specifically, consider the Hamiltonian
\begin{equation}
H = -J \left( \sum_{j=1}^{N}  \sigma_j^z \sigma_{j+1}^z + g \sum_{j=1}^{N} \sigma_j^x\right),
\label{eq:ising_hamiltonian}
\end{equation}
where $\sigma_j^{x,y,z}$ are Pauli operators acting on site $j$. This Hamiltonian has $\mathbb{Z}_2$ symmetry $\prod_j \sigma_j^x$. This Hamiltonian is the paradigmatic example of quantum phase transition in one-dimension \cite{sachdev2011qpt}: the product state Eq.~\eqref{eq:product} and the GHZ state Eq.~\eqref{eq:GHZ} are the ground states of the Hamiltonian at $g \rightarrow \infty$ and $g = 0$, respectively, where the $\mathbb{Z}_2$ symmetry is enforced. We call the product state to be in the \textit{trivial} phase and the GHZ state to be in the $\mathbb{Z}_2$-\textit{spontaneous symmetry-breaking} (SSB) phase, and accordingly, states that are connected to these two representative states by FDLU will be called the trivial states and the SSB states, respectively. The ground states for $0\leq g<1$ are in the SSB phase, the ground states for $g>1$ are in the trivial phase, and the phase transition happens at $g=1$. Our main goal is to differentiate the two phases using machine learning models and identify the phase transition. Although any state in the two phases can be chosen to be the representative state, we choose the GHZ state and product state because they are the simplest representative states and can be easily prepared on a quantum computer through an adaptive protocol \cite{Lu_2022}. 

Therefore, we propose that the train dataset is generated by applying Haar random evolution on the representative states. Specifically, we follow the four steps below:
\begin{enumerate}
    \item Choose a representative quantum state of the phase, i.e., $|1 \dots 1 \rangle$ for the trivial phase, or the GHZ state for the $\mathbb{Z}_2$-symmetry broken phase. Apply an on-site random unitary operation at each site to remove the bias towards a specific basis.
    \item Apply a random finite-depth local unitary circuit to this state. This circuit consists of a sequence of local (e.g., two-qubit) Haar random unitary gates applied in a brick-wall pattern, as illustrated in Figure~\ref{fig:haar}(a). The depth of the circuit determines how much entanglement is generated locally, while ensuring the state remains within its original phase.
    \item Perform $n_s$ rounds of random measurements on a patch of $l$ sites on the evolved state. The shadow data, i.e., the measurement basis and measurement outcomes, is represented by 3 Euler angles and 1 bit for each site. This serves as the input data for our machine learning model, as illustrated in Figure~\ref{fig:haar}(b).
    \item Repeat Step 1-3 multiple times with different random unitary circuits to create a diverse dataset for each phase. Denote the number of different circuits to be $n_b$, the dimension of the input data for a given phase is then $n_b \times n_s \times l \times 4$.
\end{enumerate}

The engineered dataset offers significant advantages: it can be scaled to arbitrary size and requires no prior knowledge of Hamiltonians or ground states of them. This independence makes its application to real physical systems highly nontrivial yet promising. We will validate our model against established physical Hamiltonians, including the Ising model in Eq.~\eqref{eq:ising_hamiltonian} and a generalization of it, ANNNI model,  to demonstrate its capabilities in Section~\ref{sec:ising}. In principle, our approach can analyze any quantum state—whether derived from a translation-invariant or non-translation-invariant Hamiltonian's ground state, or from systems subject to disorder \cite{2023PhRvX..13c1016M,2025PhRvX..15b1062M}—provided the fundamental definition of quantum phase remains valid. 

We will focus on the classification of trivial phase versus SSB phase in one-dimensional qubit systems, aiming to develop a machine learning model to classify quantum states in these phases from the shadow data on a patch.

\section{Time-series models for phase classification}\label{sec:nn}

There is a natural compatibility between shadow data and time-series machine learning models. We may think of each shadow measurement as a sequence of measurements in a random basis, creating an inherently sequential structure of the measurement outcomes that collectively encode the quantum state's properties. As a toy illustration, consider the $Z$-basis measurement on each spin of the product state in Eq.~\eqref{eq:product} versus the GHZ state in Eq.~\eqref{eq:GHZ}. The measurement outcomes of the product state are statistically independent across different spins, generating a completely random sequence consisting of 0 and 1. However, for the GHZ state, the measurement outcomes are correlated, giving a sequence of either all 0's or all 1's. 

More generally, universal features of the quantum phases, including topological entanglement entropy and its multipartite analogues, are encoded in the correlations of the measurement outcomes. Treating the shadow data as a time series, we anticipate that these universal features can be efficiently captured by modern time-series machine learning models. In this section, we will design a neural network to distinguish SSB states versus trivial states with a recurrent neural network (RNN)-based classifier to illustrate the proposal.

\begin{figure}[!htbp]
    \centering
    \includegraphics[width = \textwidth]{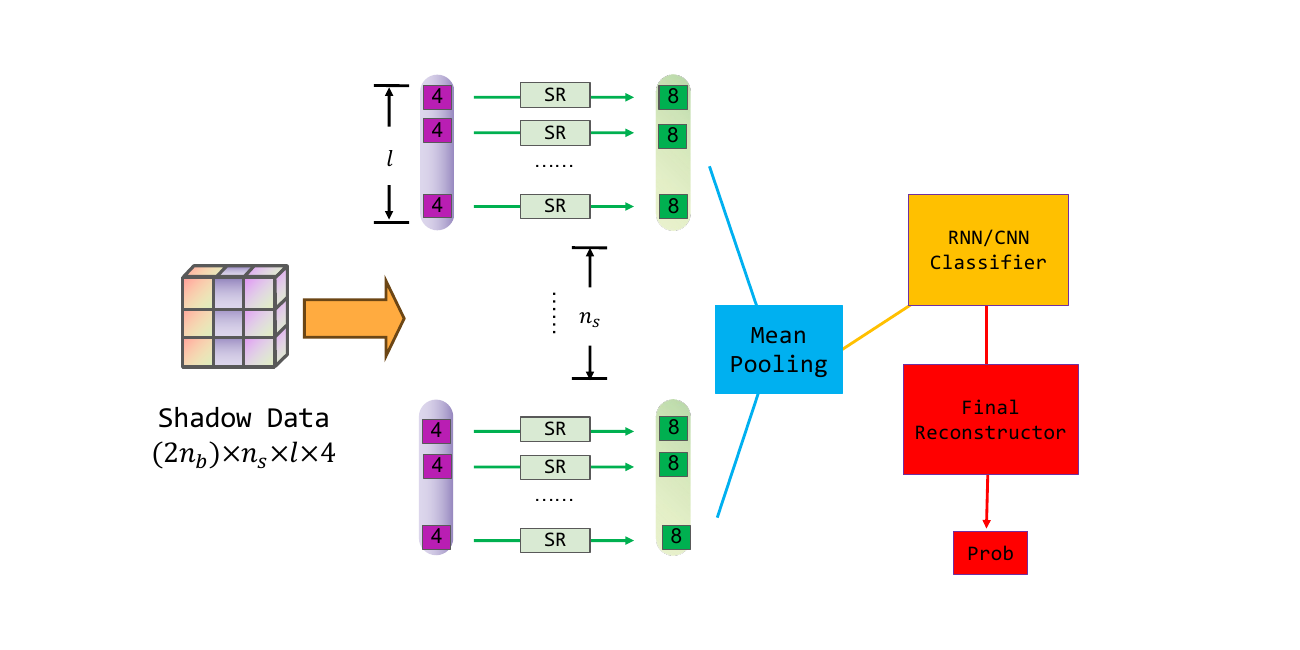}
    \caption{
    \textbf{Proposed Neural Network Architecture for Quantum Phase Classification.} This diagram illustrates the main components of our neural network. (1) The \textbf{Input} (purple block) represents the quantum states as a $(2n_b)\times n_s \times l \times 4$-dimensional tensor, representing number of quantum states ($2n_b$), number of shadows ($n_s$), patch length ($l$), and input dimension (4) consisting of Euler angles and measurement outcomes. (2) The \textbf{Shadow Reconstructor} (light green block) is a feed-forward neural network that processes each shadow independently, mapping every 4-dimensional shadow to an 8-dimensional output. (3) The \textbf{Mean Pooling} (blue block) component then computes the average of the resulting matrices over the $n_s$ different shadows. (4) The averaged representations are fed into a \textbf{Time-Series Model} (yellow block), which can be either a bidirectional recurrent neural network (BiRNN) or a convolutional neural network (CNN), designed to capture sequential dependencies or long-range correlations among different spins. (5) Finally, the output from the time-series model is passed to a \textbf{Final Reconstructor} (red block), a feed-forward network that estimates the probability of the quantum state being in a trivial (0) or SSB (1) phase.
    }
 \label{fig:neural_network}
\end{figure}

Specifically, we propose a hybrid neural network architecture that combines feed-forward layers with a bidirectional recurrent neural network (BiRNN) for quantum phase classification. The complete architecture is displayed in Figure~\ref{fig:neural_network}.

We now describe each individual block in Figure~\ref{fig:neural_network} in  detail, thereby providing a complete description of our architecture.

\begin{enumerate}
    \item \textbf{Input} (purple block): 
    
    The input quantum states are represented by a $(2n_b)\times n_s \times l \times 4$-dimensional tensor, where:
    \begin{itemize}
        \item $2n_b$: total number of quantum states in the dataset. We always construct balanced training and validation datasets, with $n_b$ trivial states and $n_b$ SSB states each.
        \item $n_s$: number of shadows for each state.
        \item $l$: length of the patch.
        \item $4$: the input dimension consisting of 3 Euler angles for the random measurement basis and 1 measurement outcome for each qubit in the patch. Specifically, we act on each site with a Haar random unitary $U(\theta, \phi, \chi)$,
\begin{equation}
    U(\theta, \phi, \chi) =
    \begin{bmatrix}
        \cos (\theta/2) e^{i(\phi+\chi)/2}  &i\sin (\theta/2) e^{i(\phi-\chi)/2} \\
        i\sin (\theta/2) e^{-i(\phi-\chi)/2} &\cos (\theta/2) e^{-i(\phi+\chi)/2}
    \end{bmatrix}
\end{equation}
with $0\leq \theta<2\pi, 0\leq \phi< 2\pi, -2\pi \leq \chi<2\pi$ and then measure the patch in $Z$ basis with output $s = 0, 1$.

    \end{itemize}

        \item \textbf{Shadow Reconstructor} (light green block):

        This component is a standard feed-forward neural network that maps 4-dimensional raw shadow measurement data to an 8-dimensional output. The output dimension is inspired by the (real) dimension of a single-qubit density matrix. It consists of several hidden linear layers, each followed by a ReLU activation, connected sequentially starting from the input layer. Importantly, the same network  processes each shadow snapshot independently. We choose the dimensions of the layers to be [4, 8, 16, 32, 64, 32, 16, 8]. The output is then a $(2n_b)\times n_s \times l \times 8$-dimensional tensor.

        \item \textbf{Mean Pooling} (blue block):

        This component computes the average of the resulting matrices over the $n_s$ different shadows and outputs a $(2n_b) \times l \times 8$-dimensional tensor. This averaging reflects the strict permutation symmetry under reshuffling of the shadows, which is reminiscent of the permutation invariant neural network or ``deep set'' architecture in \cite{2017arXiv170306114Z}.
    
        \item \textbf{Time-Series Model} (yellow block): 
        
        This component processes sequences of the 8-dimensional reconstructed representations from the Shadow Reconstructor. It uses either a BiRNN to capture sequential dependencies in both forward and reverse temporal directions, or a convolutional neural network (CNN) to capture the long-range correlation among different spins. 

        In the main text, we focus on the BiRNN. See Appendix~\ref{app:CNN} for more details of the architecture and the performance of the model if we replace the BiRNN with CNN, where a very similar performance is achieved. 

        For the BiRNN, we choose the hidden dimension to be 512, and the block outputs a $(2n_b) \times 512$-dimensional matrix. 
        
        \item \textbf{Final Reconstructor} (red block): 
        
        A final feed-forward network that takes the output from the BiRNN (or CNN) and estimates the probability of whether it is in the trivial state (0) or SSB state (1). For the BiRNN-based model, we choose the dimensions of the hidden layers and output layers to be [512, 64, 8, 1].  
\end{enumerate}

We then train our model on a concatenated dataset, where each constituent dataset has $t = 1$, $n_s =10000$ and $n_b = 2000$, but varying patch lengths ($l$) of 6, 8, 10, and 12. We evaluate the performance of the obtained model in the next section.

\section{Accuracy and test on ANNNI model}\label{sec:ising}

\subsection{States prepared using Haar random evolution}

\begin{figure}[!htbp]
\centering
\begin{subfigure}{.48\linewidth}
  \centering
  \includegraphics[width=\linewidth]{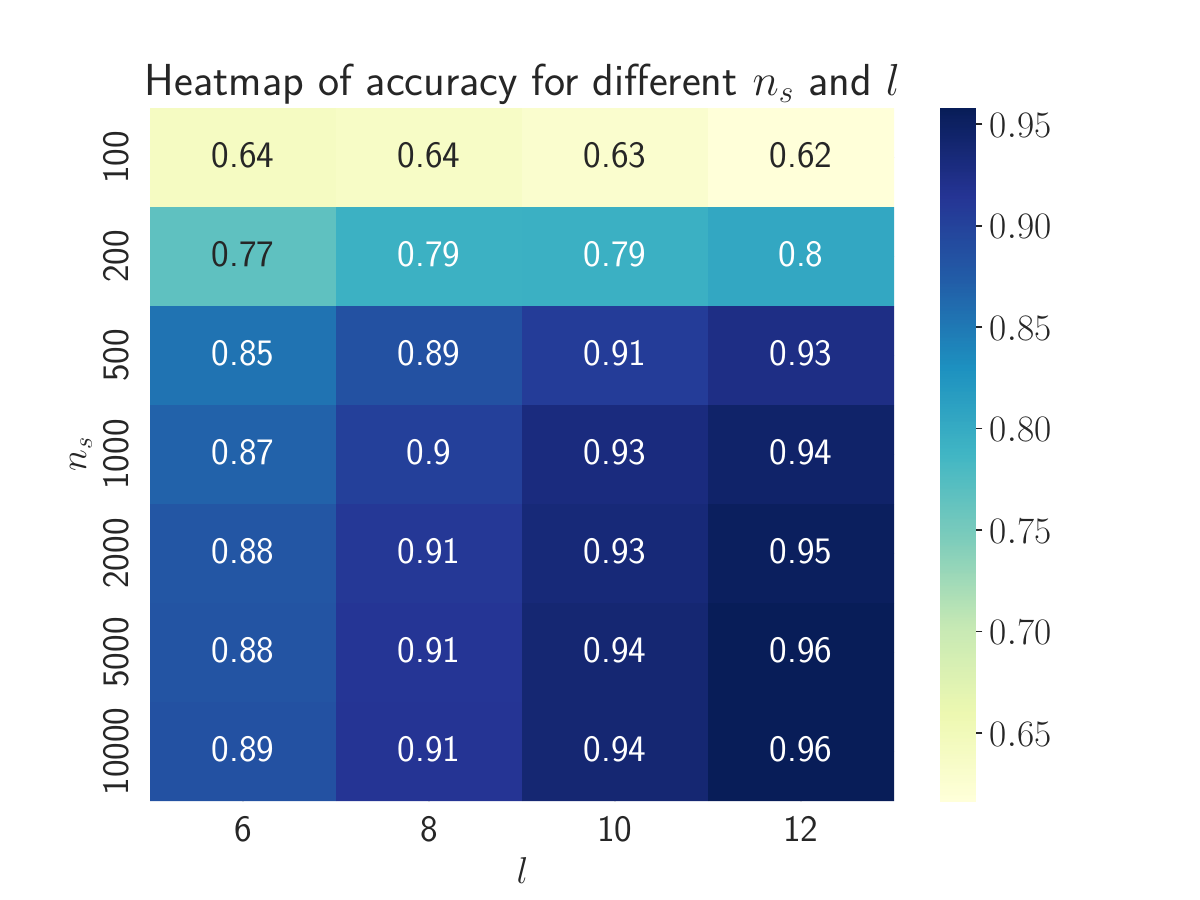}
  \caption{Classification Accuracy Heatmap}
\end{subfigure}
\hfill
\begin{subfigure}{.48\linewidth}
  \centering
  \includegraphics[width=\linewidth]{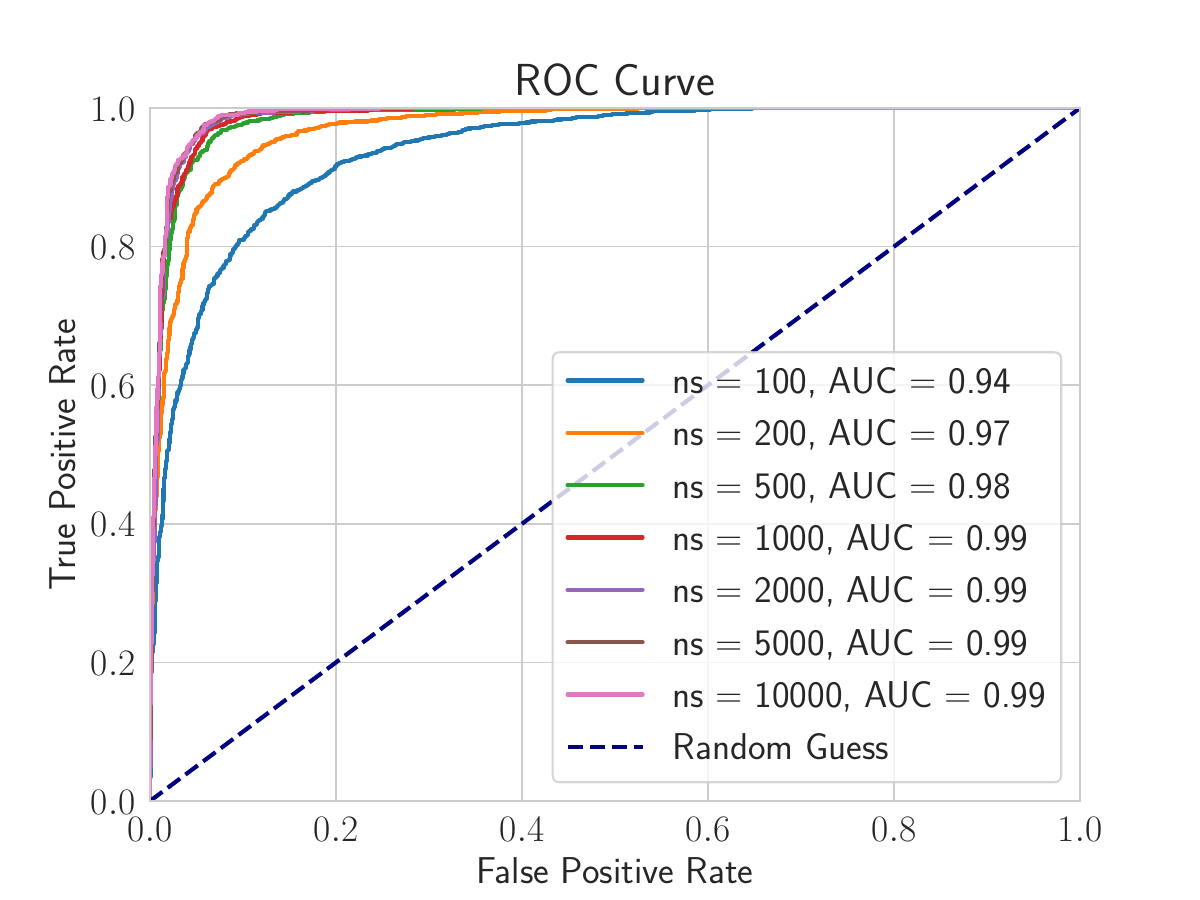}
  \caption{ROC Curve}
\end{subfigure}
\caption{
    \textbf{Performance Evaluation on Haar Random Evolved States.}
    \textbf{(a) Classification Accuracy Heatmap:} This heatmap displays the classification accuracy of our model as a function of the number of shadows ($n_s$) and the length of the patch ($l$). The evaluation is performed on a clean test set consisting of 2000 trivial states and 2000 SSB (Spontaneous Symmetry-Breaking) states, all generated via Haar random finite-depth local unitary evolution at $t=1$. The accuracy consistently improves with increasing $n_s$ and $l$, reaching an impressive 0.96 at around $n_s=5000$ shadows and a patch length of $l=12$. This demonstrates the model's ability to learn and distinguish between phases from the engineered dataset effectively.
    \textbf{(b) ROC Curve:} The Receiver Operating Characteristic (ROC) curve for the same dataset with $l=12$ and different $n_s$, which illustrates the trade-off between the True Positive Rate (sensitivity) and the False Positive Rate (1-specificity) for various classification thresholds. The Area Under the Curve (AUC) is 0.99, indicating excellent overall discrimination performance of the model. 
}
\label{fig:haar_evaluation}
\end{figure}

In order to validate the accuracy of our model, we first test the model on the states obtained from the product state and GHZ state under Haar random evolution themselves. The results are summarized in Figure~\ref{fig:haar_evaluation}.

The evaluation in Figure~\ref{fig:haar_evaluation} is performed on a clean test dataset with 2000 trivial states and 2000 SSB states. We obtain the probability of individual states being in an SSB phase and identify each state to be either trivial or SSB according to whether the probability is smaller than or bigger than 0.5, respectively. The heatmap on the left displays the classification accuracy as a function of $n_s$, the number of shadows, and $l$, the length of the patch, with the accuracy reaching 0.96 at around $n_s=5000$ and $l=12$. The Receiver Operating Characteristic (ROC) curve on the right evaluates the trade-off between the true positive rate and false positive rate for various classification thresholds, and the Area Under the Curve (AUC) is 0.99, indicating excellent overall performance. These all suggest that the model demonstrates strong classification capabilities on the engineered dataset. Additionally, we observe that the model displays reasonable accuracy (i.e. approximately 0.8) even at relatively small $n_s=200$, which is significantly better compared with protocols based on full tomography (see Appendix~\ref{app:purity}).

\subsection{Ising model and ANNNI model}

Since our model is trained on an engineered dataset, to demonstrate the effectiveness of our model in a real physical context, we will analyze our model against the Ising model and the axial next-nearest-neighbor Ising (ANNNI) model. We will see that even though the ground state of these models is nowhere in the train set, the model reproduces the phase diagram with high precision, demonstrating that the neural network learns universal features of the quantum phases.

The Hamiltonian for the axial next-nearest-neighbor Ising (ANNNI) model is given by~\cite{PhysRev.124.346,PhysRevLett.44.1502,Bak_1982}
\begin{equation}
H = -J\left( \sum_{j=1}^{N}  (\sigma_j^z \sigma_{j+1}^z - \kappa \sigma_j^z \sigma_{j+2}^z)  + g \sum_{j=1}^{N} \sigma_j^x \right),
\label{eq:annni_hamiltonian}
\end{equation}
where $\sigma_j^\alpha$ ($\alpha = x, y, z$) are Pauli matrices acting on spin-$1/2$ degrees of freedom at site $j$ of a one-dimensional lattice with $N$ sites and periodic boundary conditions. The parameter $J > 0$ is a coupling constant that sets the energy scale of the problem and does not affect the phase diagram. We associate it with the nearest-neighbor ferromagnetic exchange interaction and set $J=1$. The dimensionless coupling constants $\kappa$ and $g$ are all positive and related to the next-nearest-neighbor interaction and the transverse magnetic field, respectively. The Hamiltonian has $\mathbb{Z}_2$ symmetry generated by $\prod_j \sigma_j^x$. When $\kappa = 0$, the model reduces back to the Ising model given by Eq.~\eqref{eq:ising_hamiltonian}.

\begin{figure}
    \centering
    \includegraphics[width=0.7\linewidth]{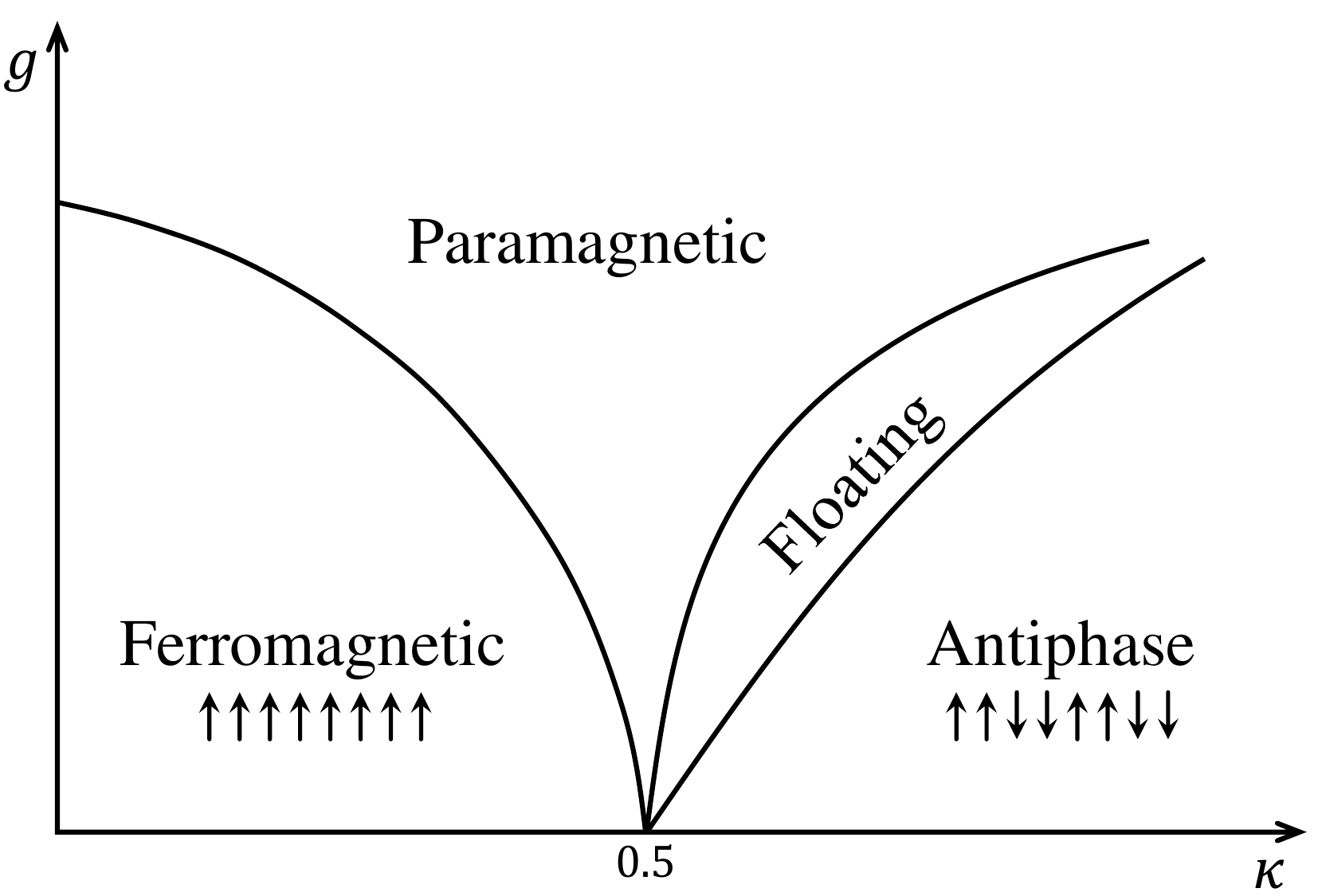}
    \caption{\textbf{Schematic phase diagram of the ANNNI model \cite{2024PhRvB.110v4422P}.}}
    \label{fig:phase_diagram_ANNNI}
\end{figure}

\begin{figure}[!htbp]
\centering
\begin{subfigure}{.48\linewidth}
  \centering
  \includegraphics[width=\linewidth]{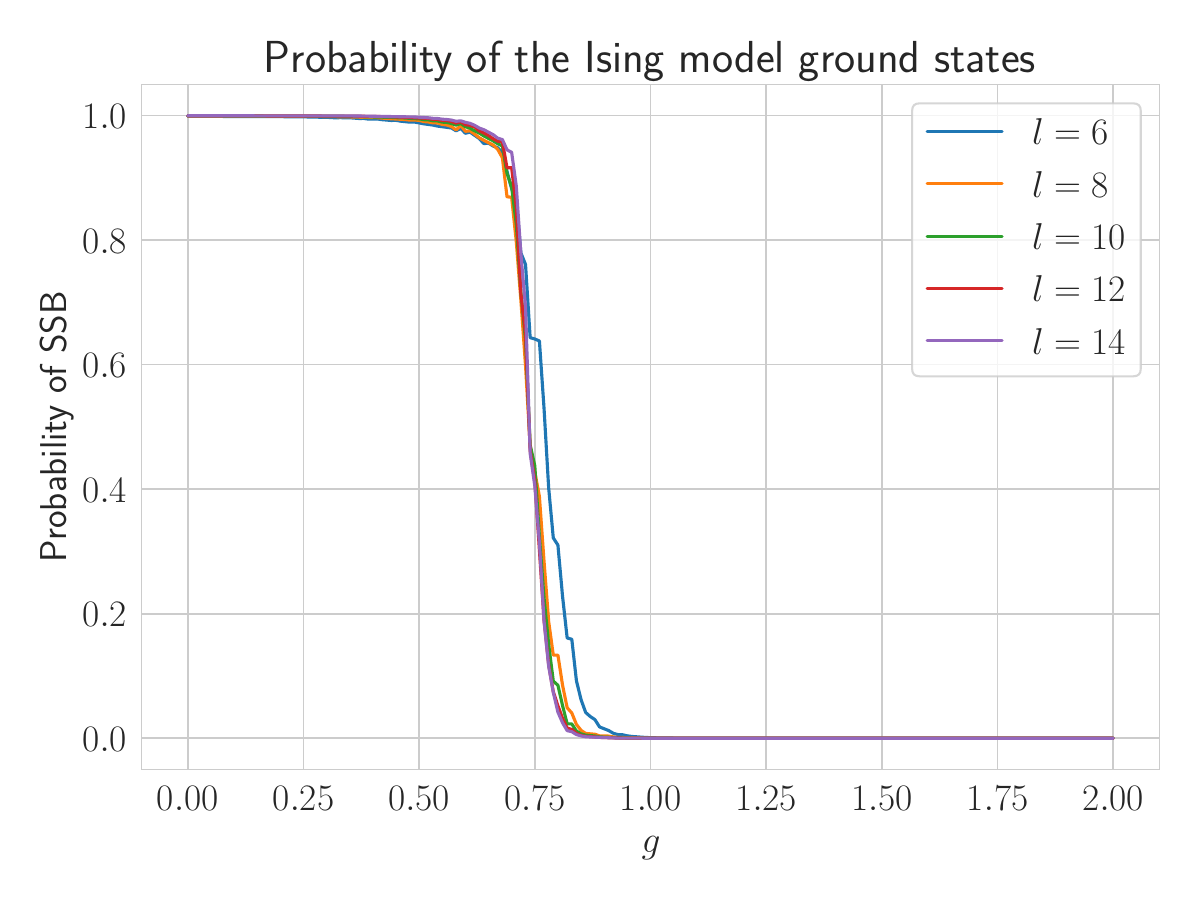}
  \caption{Ising Model Phase Classification}
\end{subfigure}
\hfill
\begin{subfigure}{.48\linewidth}
  \centering
  \includegraphics[width=\linewidth]{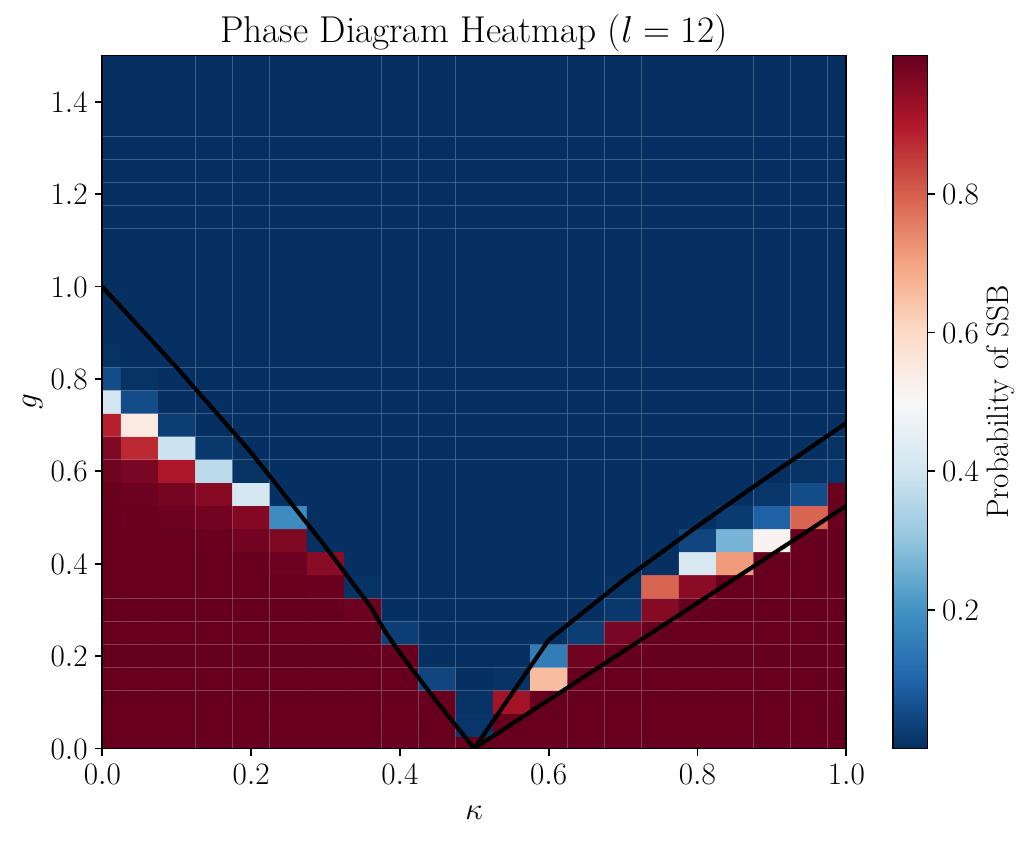} 
  \caption{ANNNI Model Phase Diagram Prediction}
\end{subfigure}
\caption{
    \textbf{Phase Diagram for Ising and ANNNI Models.}
    This
    figure displays the evaluation of our neural network model on the ground states of the Ising model and ANNNI model.
    \textbf{(a) Ising Model:} The graph displays a sharp quantum phase transition at $g=0.75$, separating SSB states on the left versus trivial states on the right. Here, $l$ is the length of the patch of the ground states we consider, and we use the shadow data with $n_s=10000$.
    \textbf{(b) ANNNI Model:} This figure presents the predicted phase diagram of the ANNNI model in the $(g, \kappa)$ parameter space, as classified by our machine learning model. Red regions represent an SSB phase, and blue regions represent a trivial phase. The black lines show the theoretical quantum phase transition lines. The shadow data for the ground states have $n_s=10000$ and $l=12$.
}
\label{fig:phase_diagram_classification}
\end{figure}

The ground-state phase diagram of the ANNNI model is displayed in Figure~\ref{fig:phase_diagram_ANNNI}. It is known to contain four distinct phases separated by three lines of quantum phase transitions: ferromagnetic, antiphase, paramagnetic, and floating phases. The paramagnetic phase possesses a unique ground state and is the trivial phase that can be connected with the product state by FDLU. If we impose $\mathbb{Z}_2$ symmetry on the ground states, the ground states in both ferromagnetic phase and antiphase are connected to GHZ state by FDLU, and hence both correspond to the $\mathbb{Z}_2$-SSB state.\footnote{The ferromagnetic phase is characterized by uniform spontaneous magnetization, with one of the ground states given by $|{11111111}\rangle$. In contrast, the antiphase breaks translational symmetry and exhibits long-range order with a four-site periodicity of the form $|{11001100}\rangle$. For our purpose, we do not impose translational symmetry on the states we consider; hence, this difference is immaterial to us. Indeed, the ground states in the ferromagnetic phase and antiphase can be connected by FDLU that breaks translational symmetry.} 
Finally, the floating phase is an intermediate region between the paramagnetic phase and the antiphase, whose precise identification is beyond the scope of the current work.\footnote{The floating phase is gapless with power-law decaying long-range correlations, in contrast to the other phases that have a finite energy gap. We do see signatures of these floating phase in our machine learning models in e.g., Figure~\ref{fig:phase_diagram_classification}, even though theoretically these gapless phases cannot be identified through a finite patch size $l$ . It is interesting to see whether/how our machine learning framework can be extended to identify these gapless phases with infinite correlation length, possibly with (shadow) information in a patch that scales with system size.} We give a more detailed description of the phase transitions in the ANNNI model in Appendix~\ref{app:annni}.

Our model successfully classifies the ground states and reproduces the known phase boundaries with high fidelity, as shown in Figure~\ref{fig:phase_diagram_classification}. In Figure~\ref{fig:phase_diagram_classification}(a), we focus on the ground states of the Ising model at $\kappa=0$. We correctly identify a sharp phase transition for the Ising model, although the phase transition point is at $g=0.75$ instead of the theoretical value at $g=1$. This result confirms  that our model has successfully captured the essential features of the Ising phase transition.\footnote{We will comment on how to improve this transition point in an updated version of the draft.}

Figure~\ref{fig:phase_diagram_classification}(b) shows the classification of the ground states for the ANNNI model. Here, red regions represent an SSB phase and blue regions represent a trivial phase. The lighter-colored regions correspond to states with a probability close to 0.5, suggesting the presence of a floating phase or an intermediate region. The black lines in the figure show the theoretical quantum phase transition lines.

The results in Figure~\ref{fig:phase_diagram_classification}(b) match the known phase diagram schematically shown in Figure~\ref{fig:phase_diagram_ANNNI} with high precision. For $\kappa < 0.5$, a clear transition line separates the blue and red regions, which aligns well with the theoretical prediction, particularly for $\kappa > 0.2$. When $\kappa > 0.5$, the model accurately identifies both the trivial and SSB states. Additionally, the presence of numerous lighter-colored blocks between the blue and red regions provides strong evidence for an intermediate floating phase. These findings offer crucial proof that our machine learning model is capable of classifying quantum phases for physical ground states it has not been trained on.

\section{Conclusion and discussion}

In this work, we introduce a data-driven framework for universal quantum phase classification that builds upon the FDLU equivalence of quantum phases, shadow tomography, and advanced time-series machine learning models. Our approach circumvents the limitations of traditional methods that rely on specific order parameters by generating training data through Haar random finite-depth local unitary evolution on a representative quantum state. As a result, we have demonstrated a systematic protocol that is independent of specific Hamiltonians and applicable to a wide range of quantum phases, potentially including those without conventional order parameters. 

The performance of our model is remarkable. On the engineered dataset of Haar-random evolved states, our framework achieves an impressive classification accuracy of 0.96 and an AUC of 0.99, demonstrating its strong capability to distinguish quantum phases purely from measurement data. More profoundly, when applied to well-known physical models like the transverse-field Ising model and the ANNNI model, our model accurately reproduces their intricate phase diagrams and critical points, despite having been trained on a Hamiltonian-agnostic dataset. This generalization capability underscores the universality and robustness of our framework, showcasing its potential for automated phase discovery in complex quantum systems where a comprehensive understanding of order parameter is lacking. 

Our machine-learning-based framework offers advantages over full tomography based methods in both efficiency and performance. The performance of our model consistently outperforms full-tomography-based method, as illustrated in Figure~\ref{fig:haar_evaluation} and Figure~\ref{fig:haar_evaluation_purity}, when $n_s$ is small. Especially, the machine learning classifier achieves very good accuracy at $n_s=500$. As such, we reduce the number of measurements and computational resources required for data acquisition compared with full-tomography-based techniques, making the method scalable and well-suited for near-term quantum devices with limited coherence time and measurement capabilities. We will investigate how our model performs under noise and other errors inherent in the current quantum computer in future work.

Our work represents a significant step towards bridging theoretical concepts of quantum phases with experimental identification on quantum computers. Future directions include extending this framework to higher-dimensional systems where genuine topological order is present \cite{2006PhRvL..96k0404K,2006PhRvL..96k0405L,2023PhRvL.131p6601K}, exploring its applicability to non-equilibrium quantum phases or those defined in mixed states \cite{2025PhRvX..15b1062M,2023PhRvX..13c1016M,lee2022symmetry,lessa2025strong,zou2023channeling,sang2024mixed,markov,zhang2025probingmixedstatephasesquantum,Lee2023quantum,chen2023separability,fan2023diagnostics,coser2019classification,de2022symmetry,sohal2024noisy}, and integrating more advanced time-series architectures like transformers to capture even richer information from shadow data. Furthermore, investigating the interpretability of the machine learning model to extract physically meaningful insights about physical quantities that differentiate different phases remains an exciting avenue for future research. Especially, from the inspiration of the machine learning model, we may be able to obtain an analytical quantity that identifies topological entanglement entropy, other anyon data \cite{2020PhRvB.101k5113K}, or symmetry-related variants of them \cite{2024arXiv241023380K,2014arXiv1410.4540B,2024PhRvX..14b1053Y} on a given lattice. Overall, this data-driven approach opens new avenues for exploring the vast landscape of quantum phases of matter.

\section*{Code availability}
The code for this project, including data generation, neural network implementation, and phase classification, will be made publicly available upon publication.

\begin{ack}
We thank Tim Hsieh, Zidu Liu, Zhi Li, and Junyu Liu for helpful discussions. 

This work was in part supported by the European Commission under the Grant Foundations of Quantum Computational Advantage, and through computational resources and services provided by Quantum Advanced Research Computing (QuARC) at the Stewart Blusson Quantum Matter Institute. We acknowledge the support of the Perimeter Institute for Theoretical Physics (PI) and the Natural Sciences and Engineering Research Council of Canada (NSERC). Research at PI is supported in part by the Government of Canada through the Department of Innovation, Science and Economic Development Canada and by the Province of Ontario through the Ministry of Colleges and Universities. 
\end{ack}

\printbibliography

\appendix

\section{CNN classifier}
\label{app:CNN}

\begin{figure}[!htbp]
\centering
\begin{subfigure}{.48\linewidth}
  \centering
  \includegraphics[width=\linewidth]{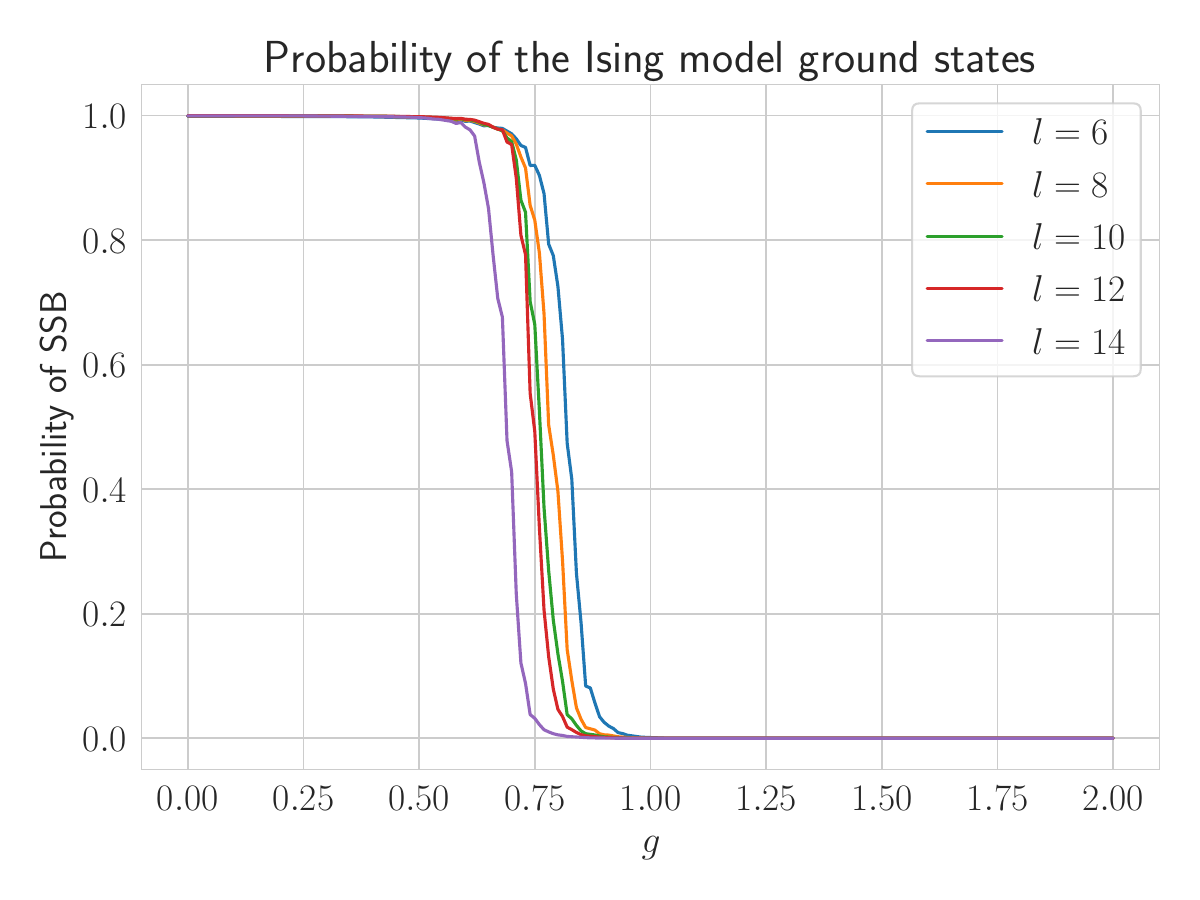}
  \caption{Ising Model Phase Classification}
\end{subfigure}
\hfill
\begin{subfigure}{.48\linewidth}
  \centering
  \includegraphics[width=\linewidth]{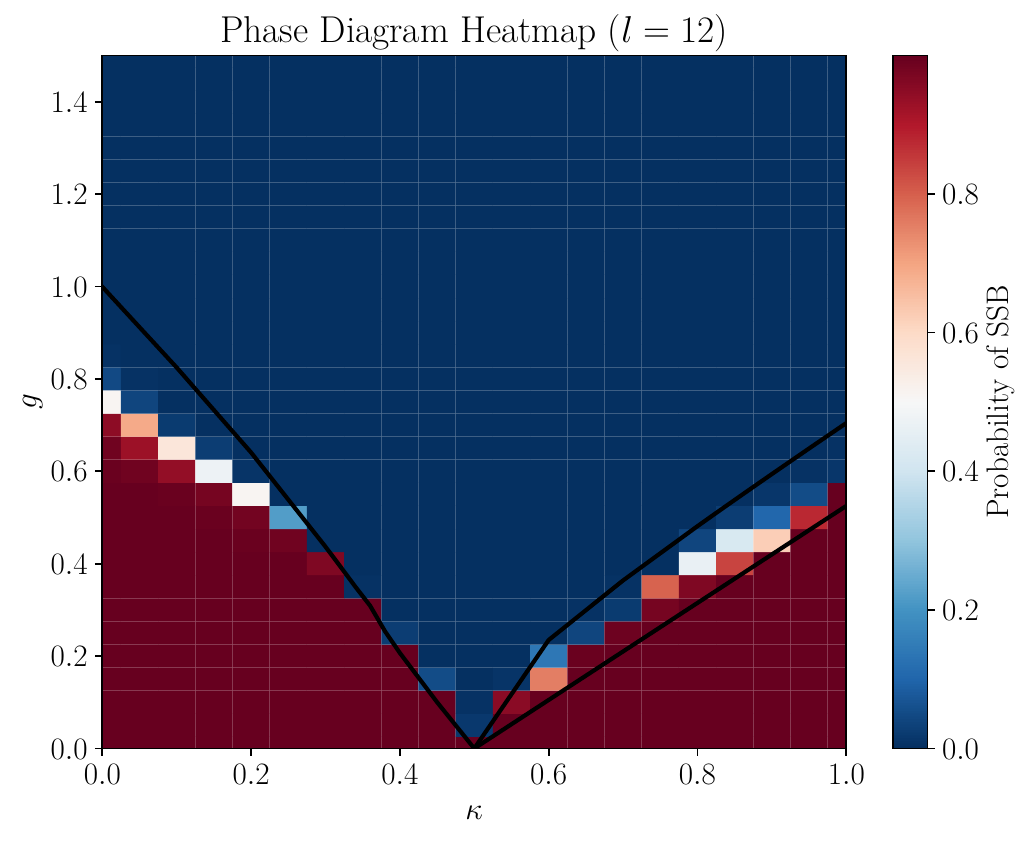} 
  \caption{ANNNI Model Phase Diagram Prediction}
\end{subfigure}
\caption{
    \textbf{Phase Diagram for Ising and ANNNI Models.} This
figure displays the evaluation of the CNN-based model on the ground states of the Ising model and ANNNI
model, using the same shadow data as Figure~\ref{fig:phase_diagram_classification}.
    \textbf{(a) Ising Model:} The graph displays a sharp quantum phase transition also at $g=0.75$, separating SSB states on the left versus trivial states on the right. 
    \textbf{(b) ANNNI Model:} This figure presents the predicted phase diagram of the ANNNI model in the $(g, \kappa)$ parameter space, as classified by our machine learning model. Red regions represent an SSB phase, and blue regions represent a trivial phase. The black lines show the theoretical quantum phase transition lines. 
}
\label{fig:phase_diagram_classification_CNN}
\end{figure}

Our framework is highly versatile, capable of accommodating a wide range of time-series models. To illustrate this flexibility, we replace the Bidirectional RNN component in Figure~\ref{fig:neural_network} with a convolutional neural network (CNN). The resulting model's performance is strikingly similar compared with the Bidirectional RNN-based model, as demonstrated in Figure~\ref{fig:phase_diagram_classification} and Figure~\ref{fig:phase_diagram_classification_CNN}. This demonstrates that the effectiveness of our framework is robust and largely independent of the specific choice of time-series model.

\begin{figure}[!htbp]
    \centering
    \includegraphics[width = \textwidth]{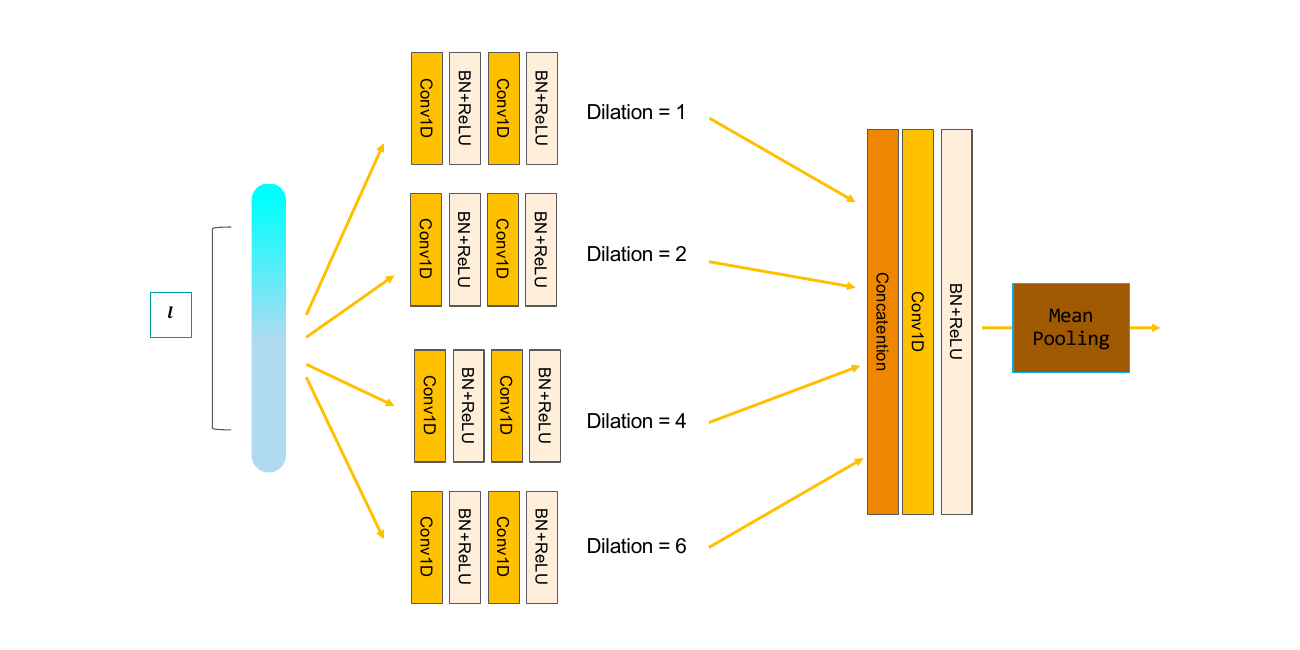}
    \caption{
    \textbf{Neural Network Architecture for CNN Classifier.} This diagram illustrates the CNN classifier component of our neural network architecture. The output matrix from the previous mean-pooling layer first undergoes four different blocks, each of which consists of two copies of 1-dimensional convolutional layer with different dilations followed by batch normalization and ReLU activation function. Then the four resulting matrices are concatenated with each other, and undergo another copy of 1-dimensional convolutional layer with no dilation. Finally, we do a mean pooling and output a matrix for the final reconstructor.
    }
 \label{fig:cnn}
\end{figure}

The \textbf{CNN Classifier} we design to replace bidirectional RNN is illustrated in Figure~\ref{fig:cnn}. The input is a $(2n_b)\times l \times 8$-dimensional tensor after the mean pooling layer. The input is then fed into four branches. Each branch consists of two copies of 1-dimensional convolutional layer followed by batch normalization and ReLU activation function. Each convolutional layer has kernel size 3 while different dilations to be equal to 1, 2, 4, and 6, respectively, and each convolutional layer also has padding equal to dilation to ensure the length to be the same before and after the layer. We choose the number of output channels to be equal to 128, and hence the output of one branch is a $(2n_b)\times l \times 128$-dimensional tensor. These varied dilations enable the CNN classifier to effectively capture correlations across different sites in the model.

These four matrices are then concatenated to form a single $(2n_b) \times l \times 512$-dimensional tensor. This is then fed into another 1-dimensional convolutional layer, followed by batch normalization and a ReLU activation function. For this layer, the kernel size is 1, with no dilation or padding, and the number of output channels is set to 128. Next, a mean pooling operation is performed over the length $l$, resulting in a $(2n_b) \times 128$-dimensional matrix. This tensor is then passed through a two-layer feedforward network with a ReLU activation function and a 0.1 dropout layer between the two layers. Finally, the output tensor is sent to the same final reconstructor, which has an identical hidden layer structure, to generate the desired probability.

We then train our model on the dataset with $t = 1$,  $n_s =10000$, $l = 12$ and $n_b = 2000$. A notable advantage of the CNN-based model is its significantly faster training time because of the intrinsic parallelizable structure that can be utilized by modern GPU.

\begin{figure}[!htbp]
\centering
\begin{subfigure}{.48\linewidth}
  \centering
  \includegraphics[width=\linewidth]{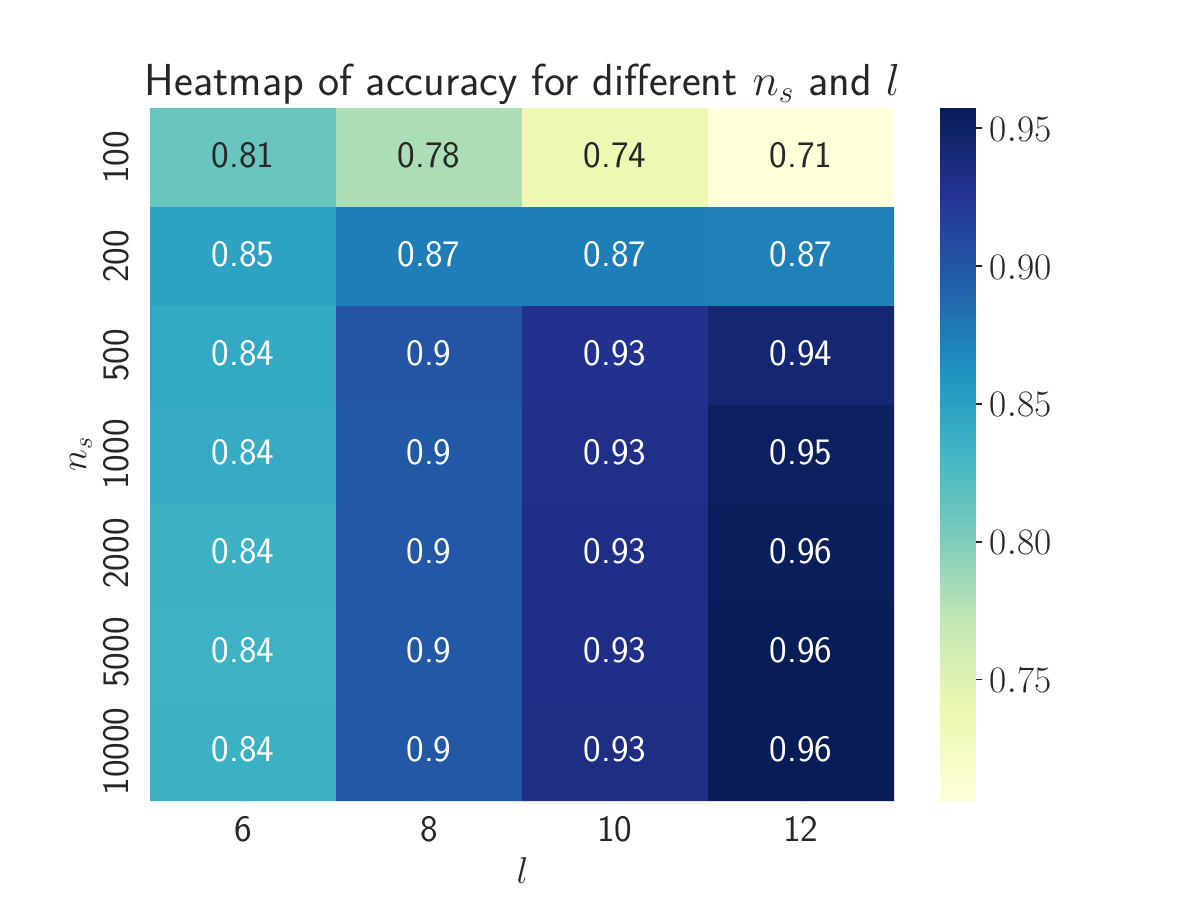}
  \caption{Classification Accuracy Heatmap}
\end{subfigure}
\hfill
\begin{subfigure}{.48\linewidth}
  \centering
  \includegraphics[width=\linewidth]{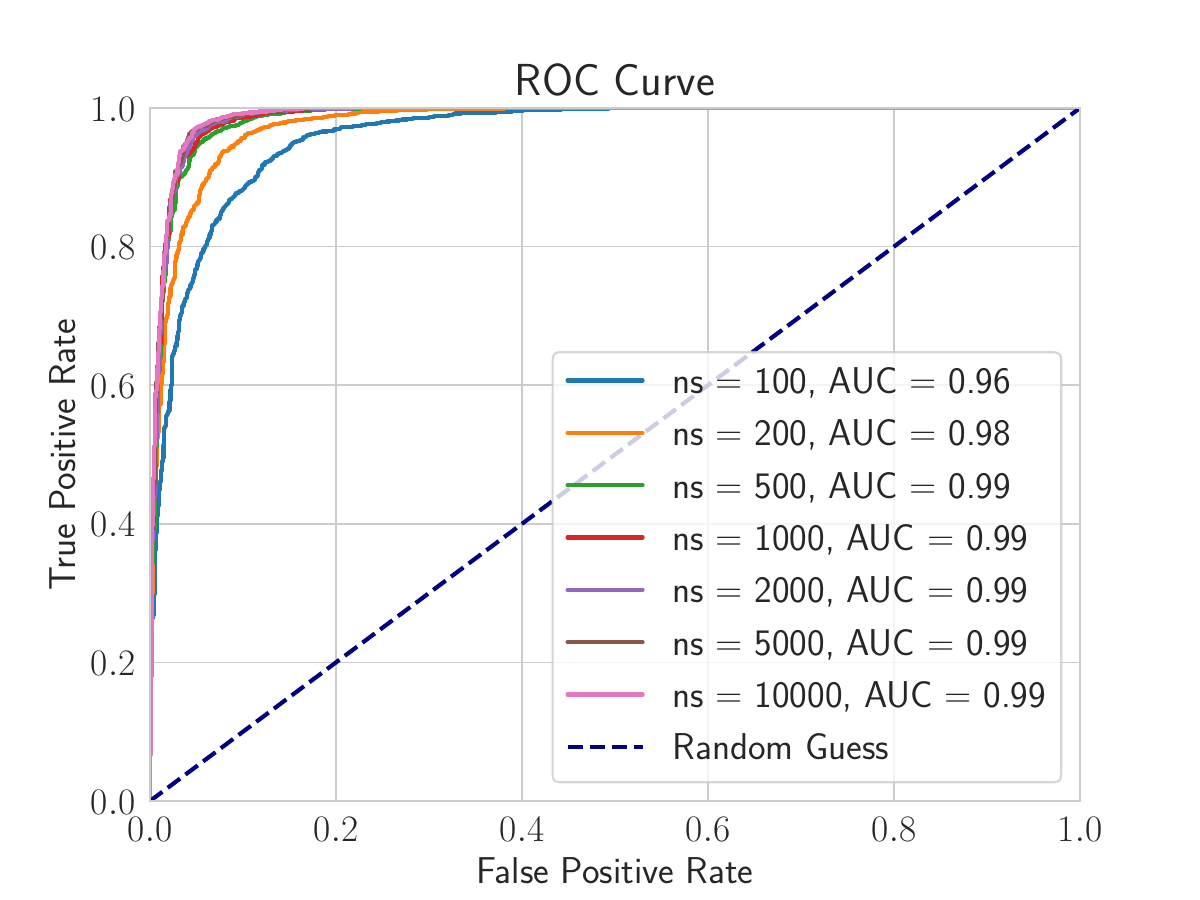}
  \caption{ROC Curve}
\end{subfigure}
\caption{
    \textbf{Performance Evaluation on Haar Random Evolved States for CNN Classifier.} This figure displays the same tests on the CNN-based model as those tests in Figure~\ref{fig:haar_evaluation}, on the same dataset consisting of 2000 trivial states and 2000 SSB states generated via Haar random FDLU at $t=1$. 
    \textbf{(a) Classification Accuracy Heatmap:} This heatmap displays the classification accuracy of our model as a function of the number of shadows ($n_s$) and the length of the patch ($l$). 
    \textbf{(b) ROC Curve:} The Receiver Operating Characteristic (ROC) curve for the dataset with $l=12$ and different $n_s$.
}
\label{fig:haar_evaluation_CNN}
\end{figure}

The performance of the obtained model is shown in Figure~\ref{fig:haar_evaluation_CNN} and \ref{fig:phase_diagram_classification_CNN}. Comparing these figures with Figure~\ref{fig:haar_evaluation} and Figure~\ref{fig:phase_diagram_classification}, we see that the performance is overall very similar. For the Haar random evolved states, the CNN-based model achieves a peak accuracy of 0.96 at around $n_s = 2000$ and $l = 12$, which exactly matches the maximum accuracy of the BiRNN-based model shown in Figure~\ref{fig:haar_evaluation}. In Figure~\ref{fig:phase_diagram_classification_CNN}(a), the CNN-based model also identifies the phase transition point at around $g=0.75$. The obtained phase diagram for the ANNNI diagram in Figure~\ref{fig:phase_diagram_classification_CNN} is strikingly similar to the phase diagram in Figure~\ref{fig:phase_diagram_classification}.

\section{Mutual information and local GEM for phase classification}
\label{app:purity}

In this appendix, we explain two non-machine-learning quantities to differentiate trivial vs. SSB states from shadow data by utilizing two entanglement-based physical quantities. The first quantity is the R\'enyi-2 mutual information of the left site and right site on a patch, reminiscent of topological entanglement entropy in higher dimensions. This quantity is directly calculable from the shadow data, yet we see that the performance is relatively not very good. The second quantity is what we call the local Geometric Entanglement Measure (GEM). This quantity requires full tomography, and achieves almost perfect accuracy at $n_s = 10000$. Still, we see that it requires a larger number of shadows compared with machine-learning-based models.

\subsection{Mutual Information}

The first entanglement-based physical quantity we use to differentiate trivial vs. SSB states is R\'enyi-2 mutual information. Here we use R\'enyi-2 mutual information instead of the von-neumann mutual information because R\'enyi-2 mutual information can be directly calculated from the shadow data.

The definition of the R\'enyi-2 mutual information for two subsystems $A$ and $B$, $I^{(2)}(A:B)$, is as follows:
\begin{equation}
    I^{{(2)}}(A:B) = S_{2}(A)+S_2(B)-S_2(AB) = \log\left[\frac{\tr (\rho_{AB})^2}{\tr (\rho_A)^2 \tr (\rho_B)^2}
    \right]
\end{equation}
where $S_2(\mathcal{O}) = -\log(\tr \rho_{\mathcal{O}}^2)$ is the R\'enyi-2 entropy of subsystem $\mathcal{O}$, and $\tr \rho_{\mathcal{O}}^2$ is the purity of the reduced density matrix $\rho_{\mathcal{O}}$. This quantity can be efficiently estimated using the median-of-means method with classical shadows~\cite{Huang_2020}. Specifically, the purity $\tr \rho_A^2$ can be estimated as:
\begin{equation}
    \begin{aligned}
\widehat{\tr \rho_A^2}(n_s, K) & =\text { median }\left\{\widehat{\tr \rho_A^2}^{(1)}(n_s, 1), \ldots, \widehat{\tr \rho_A^2}^{(K)}(n_s, 1)\right\}, \quad \text { where } \\
\widehat{\tr\rho_A^2}^{(k)}(n_s, 1) & =\frac{1}{n_s(n_s-1)} \sum_{j\neq l \in\{n_s(k-1)+1, \ldots, n_s k\}} \operatorname{tr}\left( S_A\, \hat{\rho}_j \otimes \hat{\rho}_l\right) \quad \text { for } 1 \leq k \leq K .
    \end{aligned}
\end{equation}
Here, $n_s$ is the number of shadow snapshots in each block, $K$ is the number of blocks for the median-of-means estimator, $S_A$ denotes the swap operator acting on the two-fold tensor product of subsystem $A$, and $\hat{\rho}_i$ represents a classical snapshot obtained from the shadow data. For Pauli shadows, $\hat{\rho}_i$ is reconstructed from measurement $\mathcal{M}$ as follows:
\begin{equation}
    \hat{\rho}_i = \mathcal{M}^{-1}\left[
    U_i^{\dagger}\ket{b_i}\bra{b_i}U_i
    \right]
    = \bigotimes_\alpha (3U_{i,\alpha}^{\dagger}\ket{b_{i,\alpha}}\bra{b_{i,\alpha}}U_{i,\alpha} - \mathbb{I})
\end{equation}
where $\alpha$ labels the qubits and $U_{i,\alpha}$ is the random unitary applied to qubit $\alpha$ for the $i$-th shadow, and $\ket{b_{i,\alpha}}$ is the measurement outcome of qubit $\alpha$.

Consider the R\'enyi-2 mutual information between the left-most qubit $A$ and the right-most qubit $B$ within the patch. For trivial states prepared by a Haar-random FDLU, we have $I^{(2)}(A:B) = 0$, whereas for SSB states prepared in the same way, $I^{(2)}(A:B)$ is generally nonzero. This makes $I^{(2)}(A:B)$ a natural physical quantity for distinguishing trivial from SSB states. However, it is not a perfect indicator: SSB states can have arbitrarily small $I^{(2)}(A:B)$.\footnote{For trivial states corresponding to ground states of physical Hamiltonians, $I^{(2)}(A:B)$ can also be nonzero but decays exponentially with system size.} As such, estimating $I^{(2)}(A:B)$ accurately for our classification task requires significantly more shadows than our machine-learning-based models. When the number of shadows $n_s$ is large, the computation scales as $O(n_s^2)$, making it computationally inefficient. These limitations are reminiscent of the behavior of topological entanglement entropy in higher dimensions.

\begin{figure}[!htbp]
\centering
\begin{subfigure}{.48\linewidth}
  \centering
  \includegraphics[width=\linewidth]{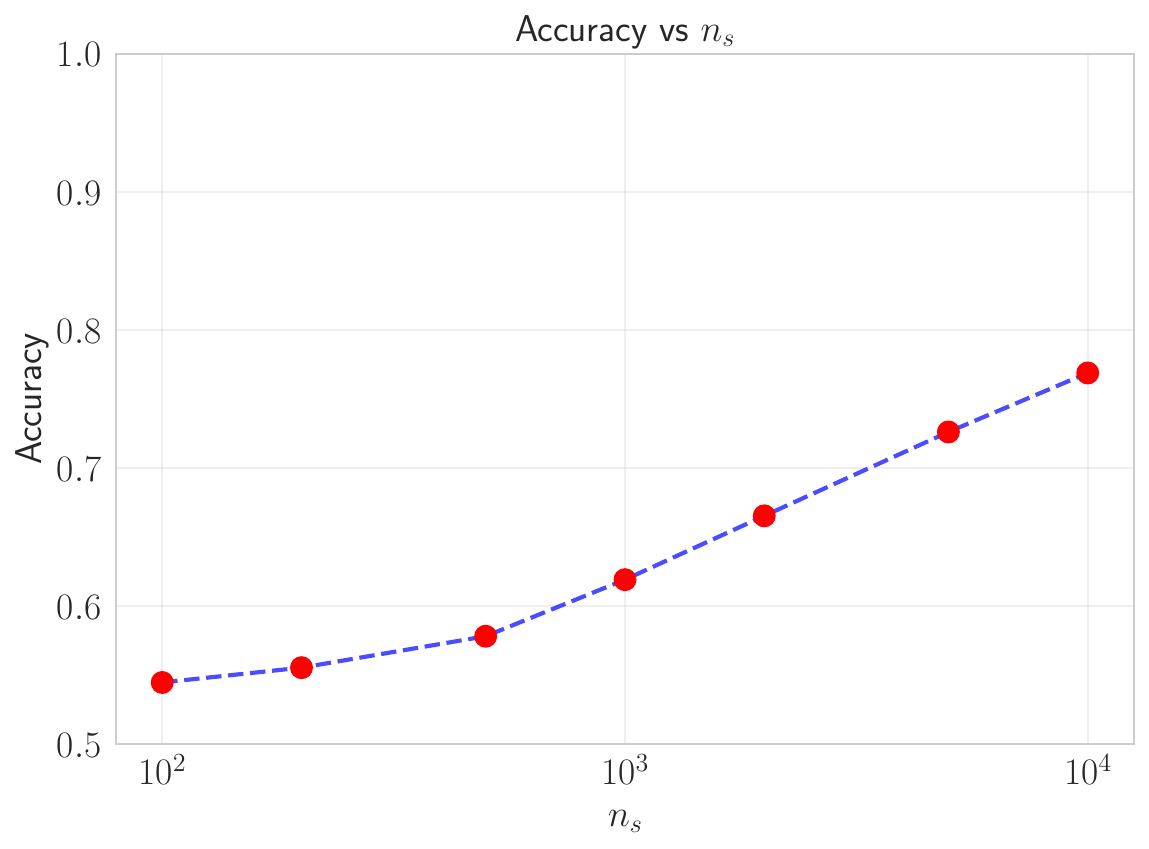}
  \caption{Classification Accuracy Heatmap}
\end{subfigure}
\hfill
\begin{subfigure}{.48\linewidth}
  \centering
  \includegraphics[width=\linewidth]{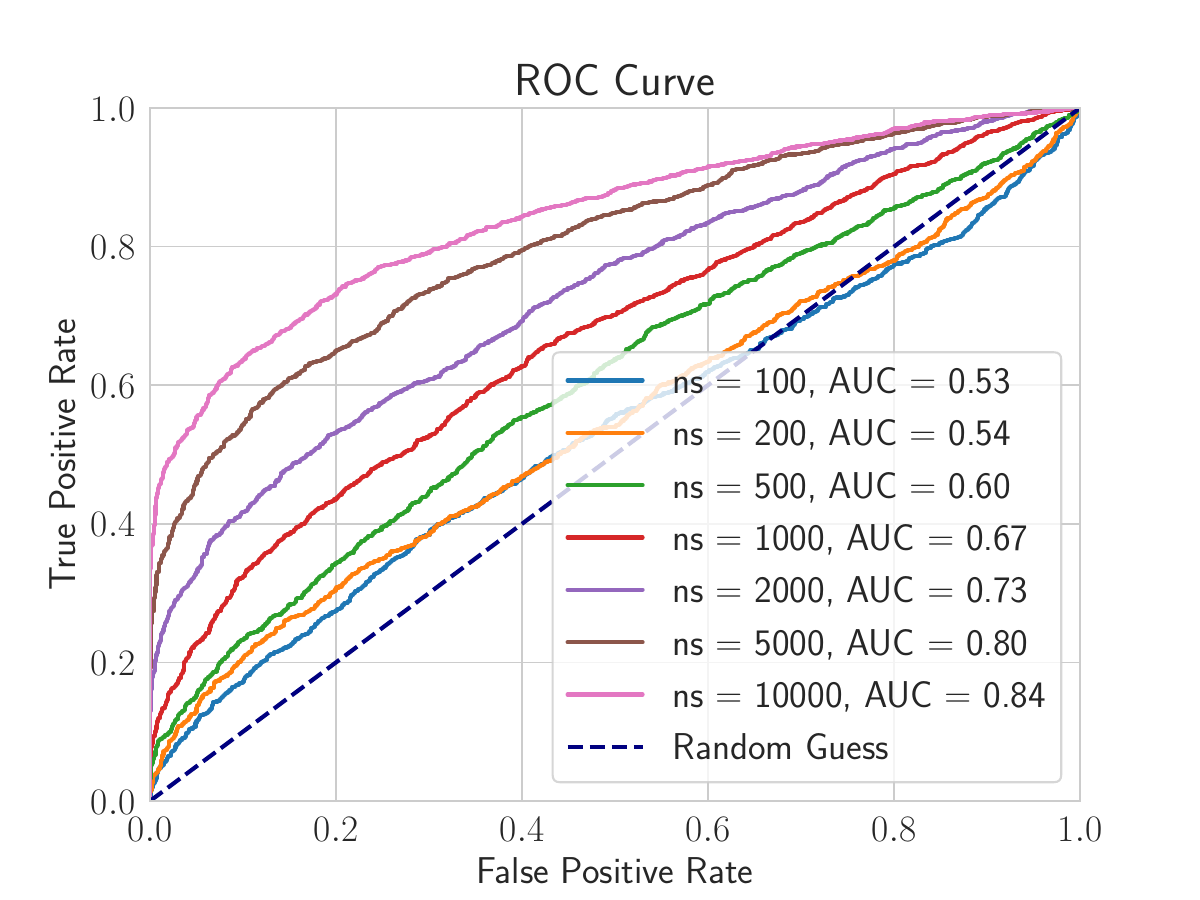}
  \caption{ROC Curve}
\end{subfigure}
\caption{
    \textbf{Performance Evaluation on Using Mutual Information as Classifier.} This figure illustrates the performance evaluation of classifying trivial versus SSB states using R\'enyi-2 mutual information, on the same dataset as those used in Figure~\ref{fig:haar_evaluation} and \ref{fig:haar_evaluation_CNN} which consists of 2000 trivial states and 2000 SSB states at $t=1$. For a given $n_s$, we select the threshold that distinguishes between trivial and SSB states to maximize accuracy, though this approach relies on information from all the other states and is thus not ideal.
    \textbf{(a) Accuracy vs $n_s$ Plot:} This plot illustrates how accuracy changes with increasing $n_s$. 
    \textbf{(b) ROC Curve:} The Receiver Operating Characteristic (ROC) curve for the dataset with different $n_s$. 
}
\label{fig:haar_evaluation_MI}
\end{figure}

The performance of mutual information given different $n_s$ is shown in Figure~\ref{fig:haar_evaluation_MI}. One important issue is that the choice of cutoff separating the two phases is ambiguous for different values of $n_s$.  We need to choose a cutoff $\epsilon$ such that when $I^{(2)}(A:B)>\epsilon$, we identify the state as SSB, otherwise we identify the state as trivial. In our analysis, we select the cutoff for each $n_s$ so as to maximize the classification accuracy. This approach is not ideal, as it uses all available information from the dataset and, in principle, the optimal cutoff can vary from one dataset to another. Even under these more favorable conditions, however, the resulting performance remains suboptimal. At $n_s=10000$, the accuracy only reaches 0.77 and the AUC is only 0.84.

\subsection{Local GEM}

Nontrivial quantum phases of matter are defined through states which cannot be connected to a product state by finite-depth local unitaries. Following this definition, Ref.~\cite{li2025entanglementneededtopologicalcodes} defines a quantity known as geometric entanglement measure (GEM) for an $N$-qubit pure quantum state $|\psi\rangle$,
\begin{equation}
    E_t(|\psi\rangle) = -\log \max_{U:\mathrm{depth ~}t} |\langle 0^{\otimes N}|U|\psi\rangle|^2
\end{equation}
where $U$ is a depth $t$ variational circuit.
For a one-dimensional lattice, the unitary is conventionally specified through
\begin{equation}
    U = \prod_{i=1}^{2t} U_i
\end{equation}
where $U_i = \otimes_x U^{[2x-1, 2x]}_i$ for odd $i$ and $U_i = \otimes_x U^{[2x, 2x+1]}_i$ for even $i$. The GEM being nonzero for all $t=O(\mathrm{polylog} ~N)$ indicates nontrivial quantum phases of matter. 

\begin{figure}
    \centering
    \includegraphics[width=0.3\linewidth]{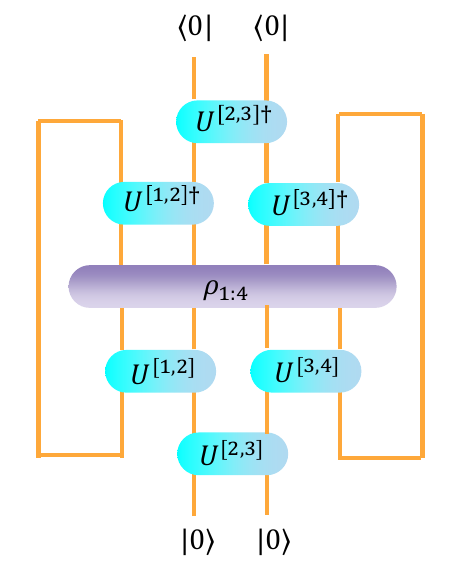}
    \caption{\textbf{The tensor network for computing the local GEM with $l=4,t=1$.} $L_1(\rho_{1:4})$ is given by minus logarithm of the maximum of the tensor network over the three unitaries $U^{[1,2]},U^{[3,4]},U^{[2,3]}$.}
    \label{fig:GEM}
\end{figure}

In this work, we distinguish SSB versus trivial states using shadows from a local patch of length $l$. Thus, we only have access to a reduced density matrix $\rho_{1:l}$ of length $l$. This motivates us to define a local analogue of GEM $L_t(\rho_{1:l})$. We require $l\geq4t$ to be larger than the lightcone of the unitary. Let the qubit be labelled by $1,2,\cdots, l$. The local GEM is defined as
\begin{equation}
    L_t(\rho_{1:l}) = -\log \max_{C} \langle 0^{\otimes (l-4t+2)} |C(\rho_{1:l})|0^{\otimes (l-4t+2)}\rangle
\end{equation}
Here, $C$ is a local quantum channel defined through the depth-$t$ variational channel
\begin{equation}
    C = C_{2t} \cdot C_{2t-1} \cdots C_{1}
\end{equation}
where the action of each channel $C_i$ is a 2-qubit unitary layer followed by tracing out the two boundary qubits,
\begin{equation}
    C_i (\rho) := \mathrm{tr}_{i,l-i+1}(U_i \rho U^{\dagger}_{i})~~ (i<2t), ~~  C_{2t} (\rho) :=U_{2t} \rho U^{\dagger}_{2t}
\end{equation}
In the equation above, $\mathrm{tr}_{i,l-i+1}$ means tracing out the $i$-th and the $(l-i+1)$-th qubit, which are located on the boundary of the chain before its action. The unitary layer $U_i$ is a brick wall unitary which only acts on the remaining degrees of freedom, i.e., from sites $i$ to $l-i+1$. For an illustration of $l=4$ and $t=1$, see Fig.~\ref{fig:GEM}.

The optimization can be done with the usual gradient optimization on unitary manifolds or greedy SVD optimization in Ref.~\cite{Evenbly_2009}. For a pure state $|\psi\rangle = U|0^{\otimes N}\rangle$, if one takes the reduced density matrix on consecutive $l$ sites, then one can show that $L_t(\rho_{1:l}) = 0$. Thus, we will use this local GEM to distinguish trivial states with SSB states. 

\begin{figure}[!htbp]
\centering
\begin{subfigure}{.48\linewidth}
  \centering
  \includegraphics[width=\linewidth]{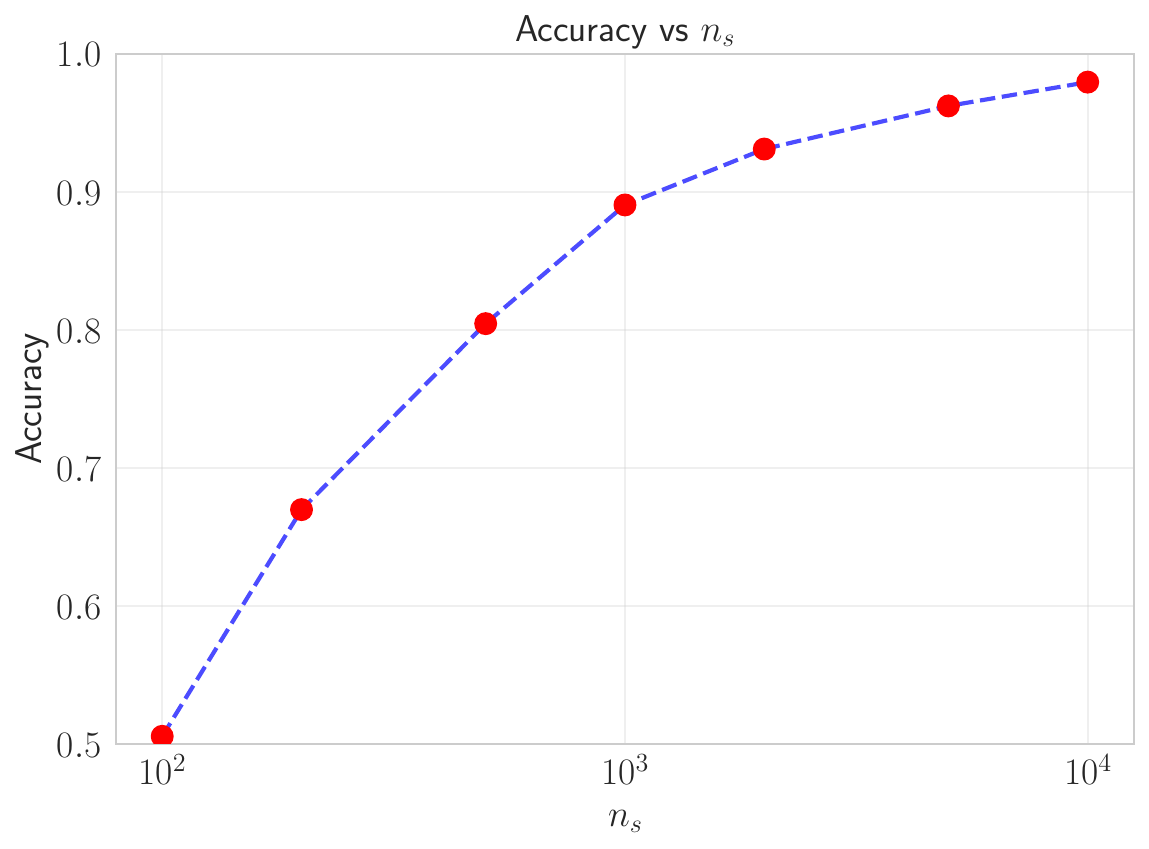}
  \caption{Classification Accuracy Heatmap}
\end{subfigure}
\hfill
\begin{subfigure}{.48\linewidth}
  \centering
  \includegraphics[width=\linewidth]{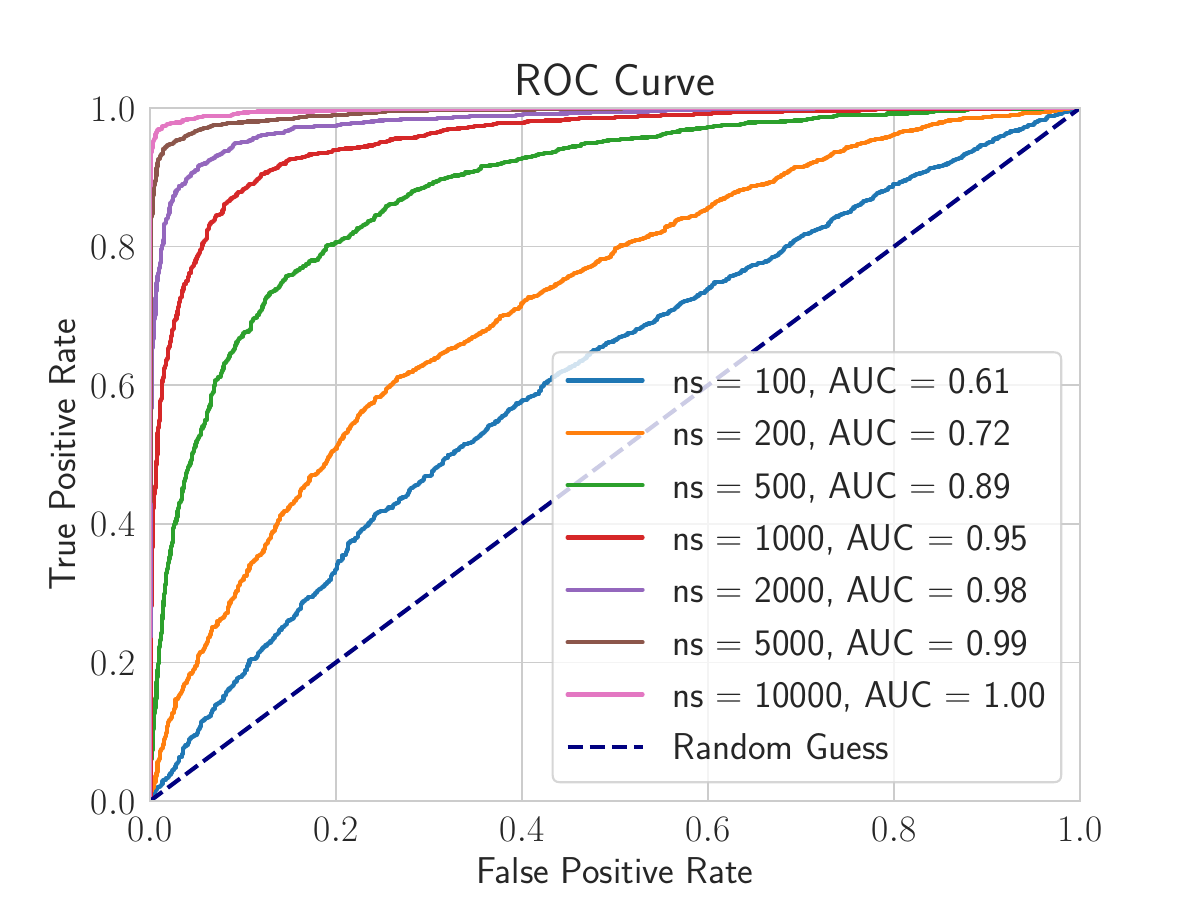}
  \caption{ROC Curve}
\end{subfigure}
\caption{
    \textbf{Performance Evaluation on Using Local GEM as Classifier.} This figure illustrates the performance evaluation of classifying trivial versus SSB states using local GEM, on the same dataset as those used in Figure~\ref{fig:haar_evaluation} and \ref{fig:haar_evaluation_CNN} which consists of 2000 trivial states and 2000 SSB states at $t=1$. For a given $n_s$, we select the threshold that distinguishes between trivial and SSB states to maximize accuracy, though this approach relies on information from all the other states and is thus not optimal.
    \textbf{(a) Accuracy vs $n_s$ Plot:} This plot illustrates how accuracy changes with increasing $n_s$. The accuracy reaches an almost perfect 0.98 at $n_s = 10000$, yet at smaller $n_s$ the full density matrix cannot be reconstructed and hence the accuracy is not as good compared with machine learning models.
    \textbf{(b) ROC Curve:} The Receiver Operating Characteristic (ROC) curve for the dataset with different $n_s$. Again, we see that it reaches the perfect 1.0, yet the performance is not very good at small $n_s$.
}
\label{fig:haar_evaluation_purity}
\end{figure}

Given the classical shadows on sites $1$ to $l$, we can reconstruct the state using
\begin{equation}
    \hat{\rho}_{1:l} = \otimes_{i=1}^l \hat{\rho}_i
\end{equation}
We then compute the local GEM $ L_t(\hat{\rho}_{1:l})$. Given a cutoff $\epsilon$, if $L_t(\hat{\rho}_{1:l})>\epsilon$, we identify the state as SSB, otherwise we identify the state as trivial. Here, just like the scenario in R\'enyi-2 mutual information, the cutoff $\epsilon$ is not pre-determined; it is chosen empirically for each dataset to achieve the best accuracy, and, in principle, can vary from dataset to dataset. We find that the best $\epsilon^{*}(n_s)$ varies significantly with $n_s$ and can be negative due to $\hat{\rho}_{1:l}$ being not positive. This feature severely limits the usage of local GEM in practical classification task as the function $\epsilon^{*}(n_s)$ is not known a priori. Under these favorable conditions, for $l=4$ and $t=1$, the result is shown in Fig.~\ref{fig:haar_evaluation_purity} for different values of $n_s$. The accuracy increases significantly with $n_s$ as one can do more accurate tomography. The accuracy can be almost perfect (0.98) at $n_s=10000$, yet not as good as the machine learning classifier for smaller $n_s$. 

\section{More details about the phase diagram of the ANNNI model}\label{app:annni}

In this appendix, we give more details about the phase diagram of the ANNNI model, with the Hamiltonian given in Eq.~\eqref{eq:annni_hamiltonian} \cite{QIP2023,2007PhRvB..76i4410B,2024PhRvA.109e2623F,SELKE1988213,2024PhRvB.110v4422P}. 

For $\kappa = 0$, the system reduces to the transverse field Ising model in Eq.~\eqref{eq:ising_hamiltonian}, which admits exact analytical solutions. In particular, there is a second-order phase transition at the critical point $g = 1$, marking the boundary between the ferromagnetic phase ($\mathbb{Z}_2$ SSB phase) for $g < 1$ and the paramagnetic phase (trivial phase) for $g > 1$.  On the other hand, when $g = 0$, the system undergoes a transition from ferromagnetic to antiphase ordering at the critical coupling $\kappa = 1/2$, even though for our purpose both phases are considered the same $\mathbb{Z}_2$-SSB phase since we ignore translation symmetries.

When both $g$ and $\kappa$ are nonzero, analytical tractability is lost, necessitating numerical approaches to map the critical boundaries. Throughout the region $0 \leq \kappa \leq 1/2$, the Ising-type transition separating paramagnetic and ferromagnetic phases extends continuously from the exactly known point $(g = 1, \kappa = 0)$ to the highly degenerate point $(g = 0, \kappa = 1/2)$, which constitutes a multicritical junction where multiple phase boundaries converge. From this multicritical point, two additional transition lines emerge and penetrate into the strongly frustrated domain where $\kappa > 1/2$. For any fixed positive transverse field $g > 0$, as the coupling strength $\kappa$ increases beyond $1/2$, the system first undergoes a Berezinskii-Kosterlitz-Thouless (BKT) transition from paramagnetic to floating phase behavior. Further increases in $\kappa$ lead to a commensurate-incommensurate (CI) transition that connects the floating phase to the antiphase regime. All these behaviors are illustrated in  Figure~\ref{fig:phase_diagram_ANNNI}.

Even though exact analytical solutions are impossible for $g, \kappa \neq 0$, approximate analytical expressions for these transition boundaries can be obtained. For instance, the critical transverse field for the Ising transition within the range $0 \leq \kappa \leq 1/2$ can be approximated as~\cite{QIP2023}
\begin{equation}
g_I(\kappa) \approx \frac{1 - \kappa}{\kappa} \left[ 1 - \sqrt{\frac{1 - 3\kappa + 4\kappa^2}{1 - \kappa}} \right].
\label{eq:ising_transition}
\end{equation}
Similarly, the critical field strength for the BKT transition in the range $1/2 < \kappa \lesssim 3/2$ is given approximately by~\cite{2007PhRvB..76i4410B}
\begin{equation}
g_{\text{BKT}}(\kappa) \approx 1.05 \sqrt{(\kappa - 0.5)(\kappa - 0.1)}.
\label{eq:bkt_transition}
\end{equation}
These analytical approximations serve as reference standards for evaluating the performance of our machine learning models.

\end{document}